\documentclass[journal]{IEEEtran}
\usepackage{cite}
\usepackage{amsmath}
\usepackage{epsfig}
\usepackage{graphicx}
\usepackage{amsmath}
\usepackage{amssymb}
\usepackage{tabularray}

\usepackage{caption}  
\usepackage{longtable} 
\usepackage{float} 
\usepackage{url}
\usepackage{xurl}
\usepackage{afterpage}
\usepackage{comment}

\usepackage{makecell} 

\usepackage{booktabs} 
\usepackage{array}
\usepackage{longtable}
\usepackage[table]{xcolor}
\usepackage[utf8]{inputenc}
\usepackage[T1]{fontenc}
\usepackage{graphicx}
\usepackage{tabularx}
\usepackage{caption}
\usepackage{adjustbox}
\usepackage{supertabular} 
\usepackage{multirow}      
\usepackage{booktabs}      
\usepackage{lipsum}        
\usepackage{hyperref}  
\usepackage{url}       
\usepackage[T1]{fontenc}
\usepackage{stfloats}

\usepackage{tikz}
\usetikzlibrary{calc}

\ifCLASSINFOpdf
\else
\fi
\begin{document}
\normalsize  
%
\title{Integrated Sensing and Communications Over the Years: An Evolution Perspective}
%
%

        \author{Di Zhang,~\IEEEmembership{Student Member,~IEEE,}
        Yuanhao Cui,~\IEEEmembership{Member,~IEEE,}
        Xiaowen Cao,~\IEEEmembership{Member,~IEEE,}
        \\Nanchi Su,~\IEEEmembership{Member,~IEEE,}
        Yi Gong,~\IEEEmembership{Member,~IEEE,}
        Fan Liu, Weijie Yuan,~\IEEEmembership{Senior Member,~IEEE,}
        \\Xiaojun Jing,~\IEEEmembership{Member,~IEEE,}
        J. Andrew Zhang,~\IEEEmembership{Senior Member,~IEEE,}
        Jie Xu,~\IEEEmembership{Fellow,~IEEE,}
        \\Christos Masouros,~\IEEEmembership{Fellow,~IEEE,}
        Dusit Niyato,~\IEEEmembership{Fellow,~IEEE,}
        and Marco Di Renzo,~\IEEEmembership{Fellow,~IEEE}

\thanks{Di Zhang, Yuanhao Cui, and Xiaojun Jing are with the School of Information and Communication Engineering, Beijing University of Posts and Telecommunications, Beijing 100876, China (e-mails: amandazhang@bupt.edu.cn; yuanhao.cui@bupt.edu.cn; jxiaojun@bupt.edu.cn). }
\thanks{Xiaowen Cao is with the College of Electronics and Information Engineering, Shenzhen University, Shenzhen 518060, China (e-mail: caoxwen@szu.edu.cn). }
\thanks{Nanchi Su is with Guangdong Provincial Key Laboratory of Aerospace Communication and Networking Technology, Harbin Institute of Technology (Shenzhen), Shenzhen 518055, China (e-mail: sunanchi@hit.edu.cn). }
\thanks{Yi Gong is with the School of Information and Communication Engineering,
Beijing Information Science and Technology University, Beijing 100192, China (e-mail: gongyi@bistu.edu.cn). }
\thanks{Fan Liu is with the National Mobile Communications Research Laboratory,
Southeast University, Nanjing 210096, China (e-mail: f.liu@ieee.org). }
\thanks{Weijie Yuan is with the School of Automation and Intelligent Manufacturing, Southern University of Science and Technology, Shenzhen 518055, China (e-mail: yuanwj@sustech.edu.cn).}
\thanks{J. Andrew Zhang is with the School of Electrical and Data Engineering, University of Technology Sydney, NSW, Australia 2007 (e-mail: Andrew.Zhang@uts.edu.au). }
\thanks{Jie Xu is with the School of Science and Engineering, the Shenzhen Future Network of Intelligence Institute (FNii-Shenzhen), and the Guangdong Provincial Key Laboratory of Future Networks of Intelligence, The Chinese University of Hong Kong (Shenzhen), Guangdong 518172, China (e-mail: xujie@cuhk.edu.cn). }
\thanks{Christos Masouros is with the Department of Electronic and Electrical Engineering,
University College London, Torrington Place, London, WC1E 7JE, UK (e-mail: c.masouros@ucl.ac.uk). }
\thanks{Dusit Niyato is with the College of Computing and Data Science, Nanyang Technological University, Singapore 639798 (e-mail: dniyato@ntu.edu.sg). }

\thanks{M. Di Renzo is with Universit\'e Paris-Saclay, CNRS, CentraleSup\'elec, Laboratoire des Signaux et Syst\`emes, 3 Rue Joliot-Curie, 91192 Gif-sur-Yvette, France (email: marco.di-renzo@universite-paris-saclay.fr), and with King's College London, Centre for Telecommunications Research -- Department of Engineering, WC2R 2LS London, United Kingdom (e-mail: marco.di\_renzo@kcl.ac.uk). }

\thanks{The work of Yuanhao Cui was supported in part by China Association for Science and Technology (CAST) Young Talent Support Program under Grant No.YESS20230663. The work of Xiaowen Cao was supported in part by the National Natural Science Foundation of China under Grant No. 62501407, the Shenzhen Science and Technology Program under Grant No. RCBS20231211090520032, and Guangdong Provincial Key Laboratory of Future Networks of Intelligence under Grant No. 2022B1212010001. The work of Yi Gong was supported in part by the Beijing Natural Science Foundation under Grant No. 4242003. The work of Weijie Yuan was supported in part by the National Natural Science Foundation of China under Grant No. 62471208. The work of Jie Xu was supported in part by the National Natural Science Foundation of China under Grants Nos. 62471424, 92267202, and U25A20390, and the Shenzhen Fundamental Research Program under Grant No. JCYJ20250604141209012. The work of Dusit Niyato was supported in part by Seatrium New Energy Laboratory, Singapore Ministry of Education (MOE) Tier 1 (RT5/23 and RG24/24). The work of M. Di Renzo was supported in part by the European Union through the Horizon Europe project COVER under Grant Agreement number 101086228, the Horizon Europe project UNITE under Grant Agreement number 101129618, the Horizon Europe project INSTINCT under Grant Agreement number 101139161, and the Horizon Europe project TWIN6G under Grant Agreement number 101182794, as well as by the Agence Nationale de la Recherche (ANR) through the France 2030 project ANR-PEPR Networks of the Future under Grant Agreement NF-YACARI 22-PEFT-0005, and by the CHIST-ERA project PASSIONATE under Grant agreements CHIST-ERA-22-WAI-04 and ANR-23-CHR4-0003-01. (Corresponding author: Yuanhao Cui.)}
}

%
%

\markboth{Journal of \LaTeX\ Class Files,~Vol.~14, No.~8, August~2015}%
{Shell \MakeLowercase{\textit{et al.}}: Bare Demo of IEEEtran.cls for IEEE Journals}
%



\maketitle
\begin{abstract}

Integrated sensing and communications (ISAC) enables efficient spectrum utilization and reduces hardware costs for beyond 5G (B5G) and 6G networks, facilitating intelligent applications that require both high-performance communication and precise sensing capabilities. This survey provides a comprehensive review of the evolution of ISAC over the years. We examine the expansion of spectrum across radio-frequency (RF) and optical ISAC, highlighting the role of advanced technologies, along with key challenges and synergies. We further discuss the advancements in network architecture from single-cell to multi-cell systems, emphasizing the integration of collaborative sensing and interference mitigation strategies. Moreover, we analyze the progress from single-modal to multi-modal sensing, with a focus on the integration of edge intelligence to enable real-time data processing, reduce latency, and enhance decision-making. Finally, we extensively review standardization efforts by 3GPP, IEEE, and ITU, examining the transition of ISAC-related technologies and their implications for the deployment of 6G networks.

\end{abstract}


%
\IEEEpeerreviewmaketitle

\section{Introduction}\label{section1}
%
%
%
%

\subsection{Background and Motivation}
Wireless communication networks are widely regarded as foundational and strategic industries, serving as the backbone for emerging sectors such as the industrial Internet of Things (IIoT) and intelligent manufacturing \cite{cui2025integratedsensingcommunicationmultifunctional}. Their continued evolution is considered a critical element in global science and technology agendas. Since the advent of radio systems in 1934, the independent development of communication and radar systems has introduced several challenges, including inefficient and competitive use of wireless resources \cite{10255711},  duplication of infrastructure and hardware \cite{zheng2019radar}, and limited cross-system interoperability \cite{8828023}. Given that electromagnetic signals inherently support both environmental sensing and data transmission, maintaining separate system designs has become increasingly redundant.


\begin{table*}
\centering
\caption{
{Comparison of Survey Focus Areas and Application Scenarios}}
\begin{tblr}{
  width = \linewidth,
  colspec = {Q[72]Q[66]Q[88]Q[90]Q[75]Q[90]Q[500]},
  row{1} = {c},
  row{2} = {c},
  cell{1}{1} = {r=2}{},
  cell{1}{2} = {c=5}{0.48\linewidth},
  cell{1}{7} = {r=2}{},
 cell{3}{1} = {c},
  cell{3}{2} = {c},
  cell{3}{3} = {c},
  cell{3}{4} = {c},
  cell{3}{5} = {c},
  cell{3}{6} = {c},
  cell{4}{1} = {c},
  cell{4}{2} = {c},
  cell{4}{3} = {c},
  cell{4}{4} = {c},
  cell{4}{5} = {c},
  cell{4}{6} = {c},
  cell{5}{1} = {c},
  cell{5}{2} = {c},
  cell{5}{3} = {c},
  cell{5}{4} = {c},
  cell{5}{5} = {c},
  cell{5}{6} = {c},
  cell{6}{1} = {c},
  cell{6}{2} = {c},
  cell{6}{3} = {c},
  cell{6}{4} = {c},
  cell{6}{5} = {c},
  cell{6}{6} = {c},
  cell{7}{1} = {c},
  cell{7}{2} = {c},
  cell{7}{3} = {c},
  cell{7}{4} = {c},
  cell{7}{5} = {c},
  cell{7}{6} = {c},
  cell{8}{1} = {c},
  cell{8}{2} = {c},
  cell{8}{3} = {c},
  cell{8}{4} = {c},
  cell{8}{5} = {c},
  cell{8}{6} = {c},
  cell{9}{1} = {c},
  cell{9}{2} = {c},
  cell{9}{3} = {c},
  cell{9}{4} = {c},
  cell{9}{5} = {c},
  cell{9}{6} = {c},
  cell{10}{1} = {c},
  cell{10}{2} = {c},
  cell{10}{3} = {c},
  cell{10}{4} = {c},
  cell{10}{5} = {c},
  cell{10}{6} = {c},
  cell{11}{1} = {c},
  cell{11}{2} = {c},
  cell{11}{3} = {c},
  cell{11}{4} = {c},
  cell{11}{5} = {c},
  cell{11}{6} = {c},
  cell{12}{1} = {c},
  cell{12}{2} = {c},
  cell{12}{3} = {c},
  cell{12}{4} = {c},
  cell{12}{5} = {c},
  cell{12}{6} = {c},
  cell{13}{1} = {c},
  cell{13}{2} = {c},
  cell{13}{3} = {c},
  cell{13}{4} = {c},
  cell{13}{5} = {c},
  cell{13}{6} = {c},
  cell{14}{1} = {c},
  cell{14}{2} = {c},
  cell{14}{3} = {c},
  cell{14}{4} = {c},
  cell{14}{5} = {c},
  cell{14}{6} = {c},
  cell{15}{1} = {c},
  cell{15}{2} = {c},
  cell{15}{3} = {c},
  cell{15}{4} = {c},
  cell{15}{5} = {c},
  cell{15}{6} = {c},
  cell{16}{1} = {c},
  cell{16}{2} = {c},
  cell{16}{3} = {c},
  cell{16}{4} = {c},
  cell{16}{5} = {c},
  cell{16}{6} = {c},
  cell{17}{1} = {c},
  cell{17}{2} = {c},
  cell{17}{3} = {c},
  cell{17}{4} = {c},
  cell{17}{5} = {c},
  cell{17}{6} = {c},
  cell{18}{1} = {c},
  cell{18}{2} = {c},
  cell{18}{3} = {c},
  cell{18}{4} = {c},
  cell{18}{5} = {c},
  cell{18}{6} = {c},
  cell{19}{1} = {c},
  cell{19}{2} = {c},
  cell{19}{3} = {c},
  cell{19}{4} = {c},
  cell{19}{5} = {c},
  cell{19}{6} = {c},
  cell{20}{1} = {c},
  cell{20}{2} = {c},
  cell{20}{3} = {c},
  cell{20}{4} = {c},
  cell{20}{5} = {c},
  cell{20}{6} = {c},
  cell{21}{1} = {c},
  cell{21}{2} = {c},
  cell{21}{3} = {c},
  cell{21}{4} = {c},
  cell{21}{5} = {c},
  cell{21}{6} = {c},
  cell{22}{1} = {c},
  cell{22}{2} = {c},
  cell{22}{3} = {c},
  cell{22}{4} = {c},
  cell{22}{5} = {c},
  cell{22}{6} = {c},
  cell{23}{1} = {c},
  cell{23}{2} = {c},
  cell{23}{3} = {c},
  cell{23}{4} = {c},
  cell{23}{5} = {c},
  cell{23}{6} = {c},
  cell{24}{1} = {c},
  cell{24}{2} = {c},
  cell{24}{3} = {c},
  cell{24}{4} = {c},
  cell{24}{5} = {c},
  cell{24}{6} = {c},
  cell{25}{1} = {c},
  cell{25}{2} = {c},
  cell{25}{3} = {c},
  cell{25}{4} = {c},
  cell{25}{5} = {c},
  cell{25}{6} = {c},
  cell{26}{1} = {c},
  cell{26}{2} = {c},
  cell{26}{3} = {c},
  cell{26}{4} = {c},
  cell{26}{5} = {c},
  cell{26}{6} = {c},
  cell{27}{1} = {c},
  cell{27}{2} = {c},
  cell{27}{3} = {c},
  cell{27}{4} = {c},
  cell{27}{5} = {c},
  cell{27}{6} = {c},
  cell{28}{1} = {c},
  cell{28}{2} = {c},
  cell{28}{3} = {c},
  cell{28}{4} = {c},
  cell{28}{5} = {c},
  cell{28}{6} = {c},
  cell{29}{1} = {c},
  cell{29}{2} = {c},
  cell{29}{3} = {c},
  cell{29}{4} = {c},
  cell{29}{5} = {c},
  cell{29}{6} = {c},
  vlines,
  hline{1,3-30} = {-}{},
  hline{2} = {2-6}{},
}
\textbf{Reference}                                                   & \textbf{Focus of Discussion}        &                                                 &                                                   &                                           &                                       & \textbf{Application Scenarios}                                                                                                                                                                                                                      \\
& {\textbf{RF to }\\\textbf{Optical}} & {\textbf{Single-Cell to }\\\textbf{Multi-Cell}} & {\textbf{Single-Modal  }\\\textbf{to Multi-Modal}} & \textbf{Security and Privacy} & \textbf{Standardization Process} &     \\

\cite{9606831} & $\times$ & $\triangle$ & $\times$ & $\times$ & $\times$ & IoT applications \\

\cite{9737357} & $\times$ & $\triangle$ & $\times$ & $\times$ & $\triangle$ & Dual-functional wireless networks in 6G \\

\cite{9585321} & $\times$ & \checkmark & $\triangle$ & $\times$ & $\times$ & Perceptive mobile network structure, signal processing \\

\cite{10614082} & $\times$ & \checkmark & $\times$ & $\times$ & $\times$ & Resource, node, and infrastructure-level cooperation \\

\cite{10418473} & $\times$ & $\triangle$ & $\times$ & $\times$ & $\triangle$ & Vehicular networks, 6G applications \\

\cite{9296833} & $\times$ & \checkmark & $\times$ & $\times$ & $\times$ & Perceptive mobile networks \\

\cite{10049304} & $\times$ & $\triangle$ & \checkmark & $\times$ & $\times$ & Multi-view sensing architectures and designs \\

\cite{10049817} & $\times$ & $\triangle$ & $\times$ & $\times$ & $\times$ & ISAC hardware testbeds and performance evaluation \\

\cite{8828030} & $\times$ & $\triangle$& $\times$ & $\times$ & $\times$ & Signal processing for ISAC \\

\cite{10.1145/3310194} & $\times$ & $\triangle$ & $\times$ & $\times$ & $\times$ & Wi-Fi / CSI-based sensing \\

\cite{10188491} & $\times$ & $\triangle$ & $\times$ & $\times$ & $\triangle$ & Historical development of ISAC \\

\cite{9705498} & $\times$ & \checkmark & $\times$ & $\times$ & $\times$ & Theoretical performance bounds \\

\cite{8999605} & $\triangle$ & $\times$ & $\times$ & $\times$ & $\times$ & Spectrum and hardware sharing \\

\cite{9540344} & $\times$ & $\triangle$ & $\times$ & $\times$ & $\times$ & Waveform design and MIMO processing \\

\cite{10004900} & $\times$ & $\triangle$ & $\times$ & $\times$ & $\times$ & UAV network structure \\

\cite{10770016} & $\times$ & $\triangle$ & $\triangle$ & $\times$ & $\times$ & Multi-carrier waveform design and learning techniques \\

\cite{10726912} & $\times$ & \checkmark & $\times$ & $\times$ & $\times$ & Multi-cell ISAC networks, interference management \\

\cite{11015430} & $\times$ & $\triangle$ & $\triangle$ & \checkmark & $\triangle$ & AI/ML-assisted ISAC architecture and secure implementation \\

\cite{10536135} & $\triangle$ & $\triangle$ & $\times$ & $\times$ & $\triangle$ & Visionary techniques for 6G ISAC \\

\cite{10781415} & $\times$ & $\times$ & $\times$ & $\times$ & $\times$ & RIS-NOMA-based ISAC techniques \\

\cite{10608156} & $\triangle$ & \checkmark & $\triangle$ & \checkmark & $\times$ & Secure architecture across Hardware/Omniscient/Application layers \\

\cite{10646523} & $\times$ & $\times$ & $\times$ & $\times$ & \checkmark & Channel modeling for communication/sensing and RCS models \\

\cite{10330577} & \checkmark & $\triangle$ & \checkmark & $\times$ & $\times$ & Unified multi-modal ISAC framework \\

\cite{10770127} & $\times$ & \checkmark & $\times$ & $\times$ & $\times$ & Interference management for multi-cell ISAC systems \\

\cite{10989512} & $\times$ & $\times$ & $\times$ & \checkmark & $\times$ & RIS-enhanced physical-layer security for ISAC \\

\cite{10833623} & \checkmark & $\triangle$ & $\times$ & $\times$ & $\times$ & RIS for multi-function ISAC \\

\textbf{This Survey} &
\checkmark & \checkmark & \checkmark & \checkmark & \checkmark &
\textbf{Emerging issues and impacts of ISAC, including the integration of the operating spectrum across RF and optical, the evolution from single-cell to multi-cell networks, the development of multi-modal cooperation, security and privacy challenges, and progress in standardization} \\

\end{tblr}
\label{tab:exsiting survey}
\caption*{\raggedright
(*) \checkmark\;\; Indicates major and systematic coverage of the corresponding dimension, typically supported by dedicated analysis or structured discussion. \\
(*) $\triangle$\;\; Indicates partial or incidental coverage, where the dimension is mentioned but not treated as a primary focus. \\
(*) $\times$\;\; Indicates that the dimension is not explicitly addressed in the survey. \\
}
\end{table*}

In response, integrated sensing and communication (ISAC) has emerged since the mid-2010s as a unified paradigm in which wireless systems are designed to simultaneously sense the physical environment and transmit information through the co-design of shared waveforms, spectrum, and hardware platforms \cite{9606831,9737357}. In this context, sensing refers broadly to radar-type functions, including object detection, tracking, and recognition, as well as behavior analysis and environmental awareness. Since its formal introduction, and particularly after 2020, significant progress has been achieved in key areas such as joint signal design, multifunctional antenna systems, spectrum sharing, and edge intelligence. Indeed, ISAC is transitioning from research to large-scale deployment, driven by both industrial adoption and standardization initiatives. In the industrial domain, Huawei has advanced ISAC deployment in vehicular networks \cite{huawei2022isac}, Ericsson has extended its application to intelligent transportation and urban infrastructure \cite{ericsson2022isac}, and Qualcomm has explored millimeter-wave ISAC for autonomous driving and IIoT applications \cite{qualcomm2022isac}. On the standardization side, 3GPP has identified 32 ISAC use cases for 6G \cite{3gpp2022isac}, ETSI’s Industry Specification Group has proposed 18 additional cases \cite{ESTI2025isac}, and IEEE finalized the first Wi-Fi sensing standard (802.11bf) in October 2024. At the global policy level, the ITU has designated ISAC as one of six core 6G application scenarios \cite{itu2021vision}, and the World Economic Forum has recognized it among the top ten emerging technologies of 2024 \cite{wef2024emergingtech}. Collectively, these developments underscore ISAC’s accelerating momentum and its potential to become a cornerstone technology in next-generation (6G) wireless networks.



\subsection{Related Survey Papers}

A broad collection of surveys has examined ISAC from complementary perspectives. Early contributions focused on spectrum and hardware reuse mechanisms \cite{8999605}, as well as waveform and multiple-input multiple-output (MIMO) techniques \cite{9540344}. These works were followed by surveys on internet of things (IoT)-oriented ISAC architectures \cite{9606831}, signal-processing techniques for 5G-Advanced/6G \cite{8828030,10012421}, channel state information (CSI)-based Wi-Fi and RF sensing \cite{10.1145/3310194}, and the historical and theoretical foundations of joint radar–communication systems \cite{10188491,9705498}, forming the foundations of ISAC research.

Built on these foundations, subsequent surveys have expanded ISAC along several technically distinct directions. One important stream has delved into physical-layer mechanisms, including multi-carrier and dual-functional waveform construction, ISAC-tailored channel modeling, and interference-management approaches for dense or heterogeneous deployments \cite{10770016,9737357,10646523,10770127}. These studies deepen the understanding of signal construction, propagation behavior, and hardware-related tradeoffs in practical ISAC deployments. At the deployment and architectural level, reconfigurable intelligent surfaces (RIS/IRS) have been widely investigated for cooperative networking \cite{10726912,10702570,10833779}, with further extensions toward RIS-assisted non-orthogonal multiple access (NOMA) for enhanced joint sensing and communication (S\&C) efficiency \cite{10781415}. Application-driven surveys have addressed ISAC in vehicular networks \cite{10418473} and uncrewed aerial vehicle (UAV)-based implementations \cite{10004900}, while the notion of perceptual mobile networks has been introduced to describe mobile infrastructures capable of dynamic S\&C fusion \cite{9296833}. Complementing analytical advances, hardware testbeds and empirical evaluations have validated ISAC concepts in practice \cite{10049817}.

Beyond physical-layer and deployment-oriented studies, another stream of literature explores ISAC’s integration with artificial intelligence (AI), security, and emerging perception paradigms. These efforts span adaptive and environment-aware transmission, secure IoT architectures, and data-centric learning frameworks for ISAC \cite{11015430,10989512,10608156}. Recent studies on multi-view sensing, generalized architectural principles, and emerging methods for multi-modal S\&C integration \cite{10330577,10049304,10536135,9585321} highlight the ongoing shift toward richer environmental awareness and learning-driven ISAC capabilities.

Viewed collectively, these surveys offer substantial depth across specific ISAC dimensions—waveforms, channel modeling, network cooperation, RIS-enhanced operation, multi-modal perception, security, and AI-assisted inference. However, their viewpoints remain inherently theme-centered: each provides a focused treatment on a subset of these dimensions, typically around a single dominant theme (e.g., security-centric, RIS-centric, or multi-modal-centric), but none attempts to synthesize how ISAC has evolved simultaneously across spectrum expansion, network-scale development, sensing-modality diversification, system-level optimization and security constraints, and progressing standardization.

As summarized in Table~\ref{tab:exsiting survey}, the existing surveys form a collection of rich but isolated slices of the broader ISAC landscape. A consolidated multi-dimensional evolutionary perspective that connects spectrum expansion, network-scale development, sensing-modality diversification, system-level optimization and standardization into a coherent trajectory is still not fully developed, even as such a perspective becomes increasingly important for guiding ISAC toward heterogeneous, scalable, and deployment-ready 6G systems.

 \begin{figure*}[!htp]
     \centering
     \includegraphics[width=0.99\linewidth]{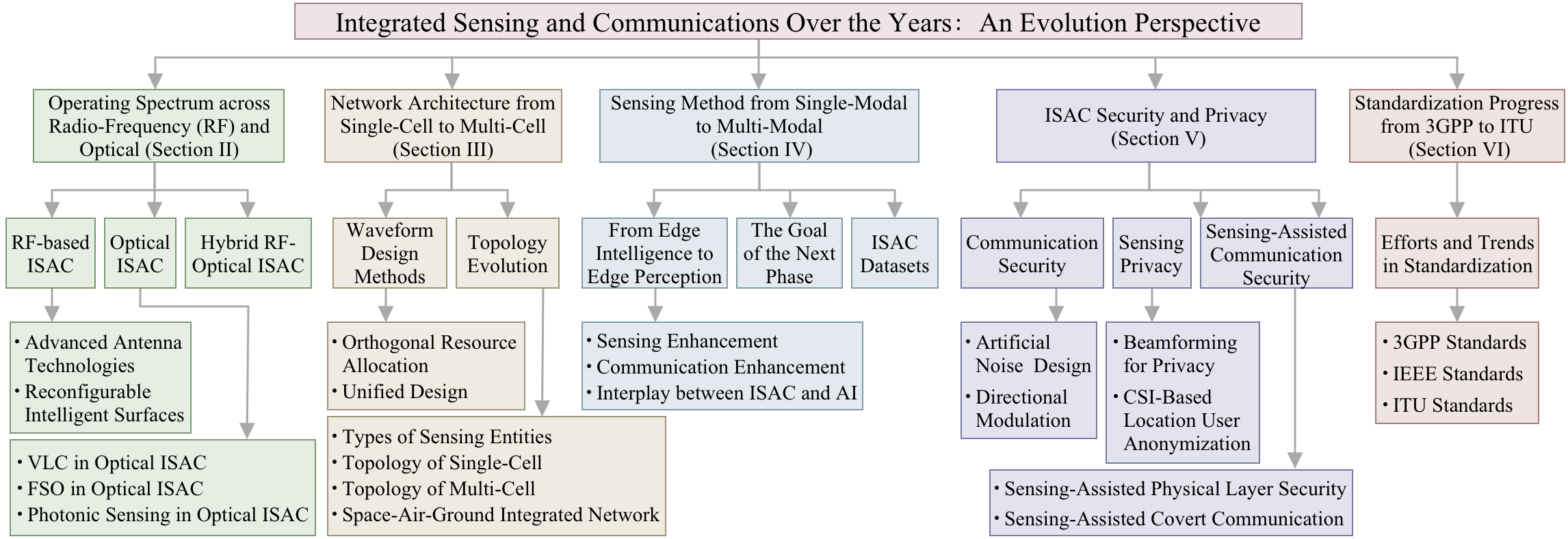}
     \caption{Structure of the Survey}
     \label{fig: paper structure}
 \end{figure*}


\subsection{Contributions}

The main contributions of this article are as follows.

\begin{itemize}
    \item Sensing from RF to Optical: We present a comparison between RF-based and optical ISAC, highlighting technical challenges and synergies and emphasizing the role of advanced antenna technologies and hybrid RF-optical architectures in enhancing 6G performance.

    \item Network Architecture from Single-Cell to Multi-Cell: We examine ISAC evolution from single-cell to multi-cell deployments, emphasizing waveform design innovations, inter-network collaboration strategies, and large-scale deployment, while addressing challenges such as synchronization, interference management, and resource allocation.

    \item Sensing Modalities from Single-Modal to Multi-Modal (Edge Perception): We trace the progression from single-modal to multi-modal sensing, with a focus on AI-enabled real-time processing, adaptive learning, and the development of high-quality datasets essential for ISAC optimization and benchmarking.
     
    \item ISAC Security and Privacy Enhancement: We investigate vulnerabilities in both communication and sensing domains, highlighting novel risks stemming from their dual-domain integration, and discuss opportunities for leveraging joint S\&C functionalities to enhance system-level security.
    
    \item Standardization Progress: We synthesize progress across major standardization bodies, including 3GPP, IEEE, and ITU, detailing advancements in New Radio (NR) and Wi-Fi sensing integration, spectrum management strategies, and emerging application domains such as low-altitude network operations.

\end{itemize}

Taken together, these contributions deliver a unified evolutionary framework that connects previously disjointed research themes into a coherent system-level narrative. This multi-dimensional synthesis, spanning spectrum, architecture, perception, security, and standardization, fills a key gap in existing ISAC surveys and lays the foundation for the broader adoption of ISAC within heterogeneous, scalable, and deployment-ready 6G systems.

The structure of this paper is outlined in Fig. \ref{fig: paper structure}. Section \ref{section2} compares RF-based and optical ISAC, detailing their respective challenges, synergies, and prospects for integrated RF-optical designs in 6G. Section \ref{section3} explores the development from single-cell toward multi-cell ISAC architectures, 
while Section \ref{section4} focuses on the evolution from single-modal to multi-modal sensing enabled by edge intelligence. 
Section \ref{section5} addresses critical security and privacy issues arising from dual-domain S\&C functionalities. 
Lastly, Section \ref{section6} reviews ongoing standardization efforts and concludes this article.


\section{Operating Spectrum across RF and Optical}\label{section2}

\subsection{RF-based ISAC}

RF technology, operating in the 3 kHz–300 GHz range, remains the foundation of wireless communication and sensing. In ISAC systems, transmitter design begins with stable carrier generation via oscillators and synthesizers, followed by modulation through mixers and upconverters to embed data or probing signals. The RF chain, comprising low-noise amplifiers (LNAs), power amplifiers (PAs), and RF filters, conditions the signal before impedance matching and directional coupling delivers it to the antenna for electromagnetic radiation. In receiver design, incident signals are captured by the antenna, amplified, downconverted, and digitized through analog-to-digital converters (ADCs) \cite{walden2002analog}, followed by digital signal processing (DSP) to recover transmitted data or extract environmental features \cite{sturm2011waveform}.

While antenna arrays, particularly massive MIMO and millimeter-wave (mmWave) configurations, enable beamforming, beam steering, and spatial multiplexing for concurrent multi-user communication and sensing, overall performance is equally constrained by RF front-end and baseband hardware. Current ISAC research emphasizes antenna architectures, suggesting a future need for deeper hardware-level performance analysis that holistically optimizes LNAs, PAs, RF filters, ADCs, and DSP modules to maximize joint S\&C efficiency.

\subsubsection{Advanced Antenna Technologies}\label{Antenna Technology}
Leveraging spatial degrees of freedom to enhance both communication and sensing, multi-antenna technologies constitute a key enabler for next-generation ISAC. Three architectural paradigms are especially impactful: centralized arrays, distributed arrays, and movable/fluid arrays.

\paragraph{Centralized Antenna Architectures}


These encompass compact and sparse array deployments. Compact arrays, with closely packed elements, offer high beamforming gain and spatial multiplexing capability, but at the cost of increased hardware complexity and a limited field of view. In contrast, sparse arrays employ non-uniform geometries (e.g., nested, coprime \cite{5609222}) to synthesize large virtual apertures, reducing the number of active RF chains while retaining beamforming functionality. Although resolution and robustness depend on array design and signal processing complexity, sparse architectures strike an effective balance between cost and angular resolution, making them attractive for ISAC hardware efficiency. For instance, Zeng et al. \cite{zeng2024fixedmovableantennatechnology} derived unified Cramér–Rao bound (CRB) and signal-to-noise ratio (SNR) scaling laws for fixed and mechanically movable uniform linear arrays (ULAs), revealing that aperture enlargement leads to a CRB that scales inversely with aperture size, while simultaneously tightening the tolerance to carrier-phase mismatch. Li et al. \cite{li2025sparse} demonstrated that uniform sparse and coprime layouts constrained to the same physical aperture as dense ULAs can preserve main-lobe characteristics and achieve competitive sum-rate and direction-of-arrival (DOA) estimation performance at low SNR, while significantly reducing the number of required RF chains. Building on these insights, Elbir et al. \cite{10371037} introduced deep learning-assisted hybrid beamforming that reduced the antenna selection search space by nine orders of magnitude while limiting spectral efficiency loss to under 5\% across 0–20 dB SNR. Taken together, these studies underscore that sparse-geometry optimization, combined with learning-based beamforming, can achieve sub-degree angular estimation accuracy with substantially reduced hardware overhead.

\begin{table*}[htp!]
\centering
\caption{Comparison of Antenna Technologies and Reconfigurable Metasurfaces in ISAC}
\label{tab:multi-antenna technologies}
\begin{tblr}{
  width=\linewidth,
  colspec={Q[90]Q[90]Q[230]Q[190]Q[170]Q[185]Q[210]Q[85]},
  row{1} = {c},
  column{2} = {c},
  cell{1}{1} = {c=2}{},
  cell{2}{1} = {r=2}{c},
  cell{2}{8} = {c},
  cell{3}{8} = {c},
  cell{4}{1} = {c=2}{c},
  cell{4}{8} = {c},
  cell{5}{1} = {c=2}{c},
  cell{5}{8} = {c},
  cell{6}{1} = {c=2}{c},
  cell{6}{8} = {c},
  vlines,
  hline{1-2,4-7} = {-}{},
  hline{3} = {2-8}{},
}
\textbf{Feature} & & \textbf{Spatial Configuration} & \textbf{Coverage} & \textbf{Hardware Cost} & \textbf{Synchronization Complexity} & \textbf{Primary Limitations}  & \textbf{Reference} \\
Centralized Antenna & {Compact\\Arrays} & Closely-packed antennas, typically uniform geometry & Limited to a localized area with narrow beams & High due to dense array elements and mutual coupling effects & Low, as elements are co-located and tightly coupled & High hardware cost or complexity, limited spatial diversity & \cite{clark2001distributed,10849704}\\
 & Sparse Arrays & Non-uniform geometry with fewer, spatially spread antennas & Wider field-of-view due to larger virtual aperture & Lower cost, requiring fewer elements and relaxed coupling & Low to moderate, depending on sparsity and processing needs & More complex signal processing for beam synthesis, with SNR trade-offs due to lower redundancy & \cite{zeng2024fixedmovableantennatechnology,li2025sparse,10371037} \\
{Distributed\\Antennas} &  & Antennas distributed across a large spatial area & Broad area coverage due to distributed diversity & Moderate to high, depending on antenna count and synchronization requirements & High, requiring advanced coordination across nodes & Requires sophisticated synchronization; scalability challenges in large networks &\cite{guo2025integrated,10464728}\\
{Movable/Fluid\\Antennas} &  & Flexible, reconfigurable positions & Enhanced coverage via dynamic repositioning & Moderate-to-high cost, depending on movement mechanisms & Moderate-to-high, contingent on control and recalibration & High energy consumption and control complexity from dynamic repositioning &\cite{ding2025movable,hao2025fluid}\\
{Reconfigurable\\Metasurfaces} &  & Arrays of passive/semi-passive elements that dynamically manipulate incident signals & Expands coverage in NLoS scenarios & Low to moderate & Low, as RIS requires no active external processing & Limited adaptability for high-mobility use cases & \cite{9140329,10858129,10833623,10738300,10841801,wang2024multi,Chen2025Integrated,11111722,di2025state}                            
\end{tblr}
\end{table*}

\paragraph{Distributed Antenna Architectures}
By geographically separating antennas and enabling cooperative operation, distributed systems offer extended spatial coverage, improved diversity, and enhanced robustness. These advantages are particularly important in multipath-rich or obstructed environments. For example, Guo et al. \cite{guo2025integrated} showed that distributed MIMO can achieve sub-meter positioning accuracy by mitigating multipath interference and exploiting spatial diversity. Similarly, Han et al.\cite{10464728} proposed a multistatic ISAC framework based on cooperative base station sensing, which significantly reduces interference compared to monostatic systems and improves positioning accuracy over both monostatic and bistatic configurations.

\paragraph{Movable/Fluid Antennas}

Movable antennas (MAs) and fluid antennas (FAs) represent a novel class of reconfigurable systems capable of adapting their physical position or geometry in real time, thereby adding new degrees of freedom to ISAC. These architectures enable dynamic beamforming through mechanical repositioning or fluid-based reconfiguration, offering significant sensing and throughput gains. In near-field ISAC, adaptive antenna positioning has been shown to yield substantial performance improvements \cite{ding2025movable}, with joint optimization of antenna placement and beamforming maximizing weighted sum rates for joint S\&C tasks. Hao et al.~\cite{hao2025fluid} demonstrated that optimizing antenna positions in a fluid-antenna–enhanced ISAC system leads to significantly higher sensing SNR (up to 50.9\% improvement) compared to traditional fixed-position antenna systems.

\subsubsection{Reconfigurable Intelligent Surfaces}

RISs have been widely used in ISAC systems to dynamically shape the electromagnetic propagation environment \cite{9140329,10858129}. Unlike conventional antennas that actively transmit or receive, RISs are composed of passive or semi-passive elements that manipulate incident waves through programmable control of their reflection and transmission coefficients, thereby enabling wireless environment reconfiguration. By adaptively adjusting the phase and amplitude of incoming signals, RISs facilitate beam steering, interference suppression, and enhanced spatial multiplexing \cite{10738300}.

In ISAC applications, RISs have been employed for coordinated beamforming and dynamic resource allocation among multiple users \cite{10841801}. More advanced implementations include stacked multi-layer transmissive RISs, which are optimized to minimize the CRB under constraints on signal-to-interference-plus-noise ratio (SINR) and transmit power \cite{wang2024multi,11111722}. Space–time-coded metasurface architectures have also been experimentally validated, supporting carrier-frequency communications and harmonic-based sensing within a single platform, which enables unified and cost-efficient operation across wireless data transmission, radar, imaging, and environmental monitoring \cite{Chen2025Integrated}.

RIS-assisted ISAC, however, faces several implementation challenges. The reconfiguration latency of RIS elements typically exceeds that of actively controlled antenna arrays. This latency limits their effectiveness in high-mobility scenarios that demand fast beam tracking. To improve adaptability and responsiveness, data-driven channel estimation methods have been proposed, reducing pilot overhead and computational complexity while maintaining accuracy \cite{9997576,10229204,10001672}. These learning-based approaches have achieved significant reduction in normalized mean square error (NMSE) compared with model-based baselines and have demonstrated improved robustness against phase noise and synchronization impairments. Nevertheless, wideband RIS operation requires highly accurate channel estimation and rigorous calibration \cite{di2025state}. Additionally, integrating into existing RF front-ends presents significant challenges, particularly in terms of impedance matching, signal routing, and system-level synchronization. These issues remain substantial barriers to large-scale deployment.

\subsubsection{Summary and Discussion}

The comparative performance of RF-based ISAC architectures is summarized in Table \ref{tab:multi-antenna technologies}. While RF-based ISAC remains the technological foundation, several limitations constrain its performance. High-frequency implementations in the W-band have achieved data rates up to 48.04 Gbps with a sensing resolution of 1.02 cm using a 16 GHz bandwidth \cite{liu2024w}. However, these systems remain affected by spectrum scarcity, multipath fading, and short channel coherence times. Notably, the fine sensing resolution is primarily enabled by the large available bandwidth rather than the carrier frequency itself. In the sub-terahertz regime around 275 GHz, sensing resolution at the sub-centimeter level has been demonstrated by fusing three 10 GHz channels \cite{10506595}, but severe atmospheric attenuation significantly reduces robustness in outdoor scenarios.

Furthermore, spectrum sharing through frequency-division multiplexing in multi-user environments introduces inter-user interference, as shown by Dong et al. \cite{dong2022demonstration}, thereby degrading the joint sensing and communication performance.

These limitations motivate the exploration of optical ISAC, which leverages abundant unlicensed spectrum and shorter wavelengths to achieve finer spatial resolution while alleviating spectrum scarcity in the RF domain.

\begin{figure*}[htp!]
    \centering
    \includegraphics[width=0.99\linewidth]{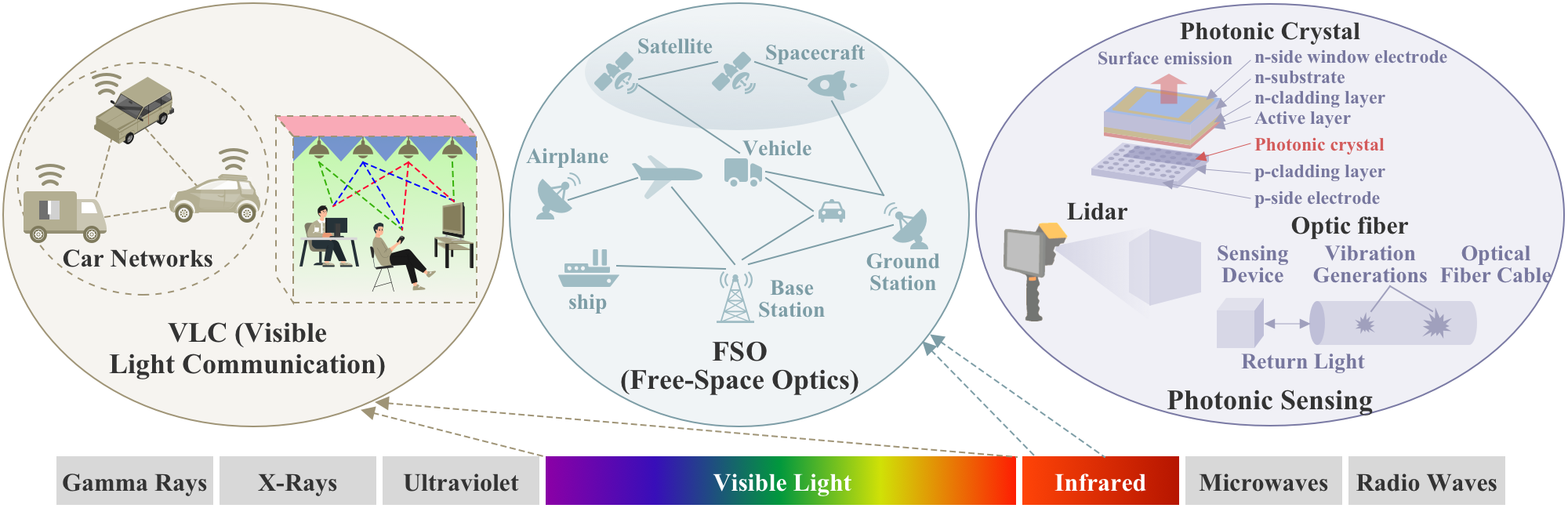}
    \caption{Optical ISAC is categorized into VLC, FSO, and Photonic Sensing}
    \label{fig:optical}
\end{figure*}

\subsection{Optical ISAC}


Optical ISAC covers multiple spectral regimes and deployment paradigms, generally categorized into three main classes: visible light communication (VLC), free-space optics (FSO), and photonic sensing, as illustrated in Fig. \ref{fig:optical}. These categories highlight the adaptability of optical modalities for joint S\&C across diverse scenarios.

\subsubsection{VLC in Optical ISAC}

VLC employs light-emitting diodes (LEDs) as transmitters and photodetectors or imaging sensors as receivers to enable simultaneous data transmission and environmental sensing. In the ISAC scenario, VLC is utilized to support centimeter-level indoor localization, motion tracking, and environmental mapping, while maintaining high data throughput via intensity-modulated optical carriers. Early foundational work by Wen et al. introduced a unified framework for optical ISAC, which integrates VLC and FSO and identifies key design challenges such as illumination-compatible waveform design and joint signal processing \cite{10574248}. Prototype systems have further demonstrated practical feasibility through LED-based dual-mode operation, combining broadcast VLC with narrow field of view (FoV) imaging for sensing, and beamformed directional links for high SNR communication\cite{10615407}. Notably, a multi-band carrierless amplitude and phase modulated VLC platform augmented with received signal strength (RSS)-based positioning was validated in a 1.2m × 1.2m × 2.16 m indoor testbed, achieving centimeter-level localization precision\cite{10559263}.

\subsubsection{FSO in Optical ISAC} 
FSO adopts modulated laser beams in the near-infrared spectrum to establish long-range, high-capacity line-of-sight (LoS) communication links, often achieving terabit-per-second data rates. Simultaneously, the narrow beam profile and interaction with the atmosphere enables ranging and velocity estimation, making FSO a compelling candidate for outdoor ISAC deployments. For instance, Khorasgani et al. presented an analytical study of capacity-distortion trade-offs in FSO-based ISAC, addressing optimal rate-distortion regions and examining the impact of nonlinear estimation techniques (e.g., maximum likelihood and maximum a posteriori) for sensing targets\cite{khorasgani2024optical}, where Bayesian CRBs were employed to derive fundamental performance limits. Moreover, Wen et al. \cite{10960484} introduced direct current-biased optical orthogonal frequency division multiplexing (OFDM)-based schemes for FSO-ISAC, modeling and balancing clipping noise and power allocation effects on spectral efficiency (communication) and Fisher information (sensing)
. Experimental work \cite{9645565} evaluated weather-induced fading and pointing loss, revealing performance degradation under fog yet stable operation in clear conditions.


\subsubsection{Photonic Sensing in Optical ISAC}

Photonic sensing technologies, including LiDAR, vibration-based fiber sensing, and other precision optical detection systems, are employed for object detection, ranging, and environmental mapping. Leveraging fiber backhaul and photonic front-ends, these systems provide high accuracy and robustness in high-frequency ISAC. Photonics-assisted W-band ISAC has demonstrated multi-target localization accuracy at the centimeter level while maintaining Gbps-class throughput via joint bandwidth–power allocation across OFDM subcarriers and FDM channels \cite{10416894}. Yan et al. \cite{yan2024w} investigated OFDM-enabled photonic ISAC with two-stage carrier frequency recovery and DSP-based correction, achieving 47.54 Gbps throughput and accurate radar detection. Additionally, quadrature phase shift keying (QPSK)-coded linear frequency modulated continuous wave (LFMCW) prototypes have achieved ranging accuracy better than 20 mm in single-target scenarios and resolution at the centimeter level in two-target sensing at 28 GHz \cite{lei2022photonics}, showing that coherent photonic ISAC can deliver high-precision sensing and high-speed communications in the lower mmWave bands.

\subsubsection{Summary and Discussion} 

Table \ref{tab:o-isac} summarizes the characteristics of optical ISAC by operating frequency, application domain, and signal transmission scheme. Despite clear advantages, optical ISAC faces significant limitations. Optical propagation is highly susceptible to environmental impairments such as fog, rain, smoke, and atmospheric turbulence. While advanced waveform designs, including LFM-continuous phase modulation (CPM) and optical OFDM, have been proposed to mitigate these effects \cite{10318167}, performance degradation under adverse weather remains a major concern, especially for outdoor and long-range deployments, limiting scalability in large or dynamic environments. Narrow beam divergence, limited field-of-view, and stringent alignment requirements further constrain widespread adoption. As shown in small-scale VLC testbeds \cite{10559263}, centimeter-level localization is feasible, but maintaining such precision in mobile or distributed deployments remains challenging; hybrid extensions via fiber backhaul or static infrastructure alleviate some issues but increase system complexity.

A further challenge is the absence of a unified framework for performance evaluation and joint optimization in optical ISAC. Unlike RF-based systems, which employ well-established metrics (such as mutual information (MI), mean square error, and ambiguity functions), there is no standardized analytical methodology for jointly assessing communication throughput and sensing fidelity in optical ISAC \cite{10419792}. This lack of unified metrics impedes the co-design of waveforms, resource allocation, and protocol layers, thereby limiting the full realization of optical-centric ISAC.

\begin{table*}[htp!]
\centering
\caption{Comparison of Optical ISAC Technology Types}
\label{tab:o-isac}
\begin{tblr}{
  width = \linewidth,
  colspec = {Q[80]Q[140]Q[150]Q[255]Q[260]},
  row{1} = {c},
  cell{2}{1} = {c},
  cell{2}{2} = {c},
  cell{3}{1} = {c},
  cell{3}{2} = {c},
  cell{4}{1} = {c},
  cell{4}{2} = {c},
  hlines,
  vlines,
}
\textbf{Feature}         & \textbf{Carrier Medium}                                         & \textbf{Communication Mode}                                    & \textbf{Sensing Mode}     & \textbf{Limitations/Challenges}  \\
VLC in ISAC              & {Visible light \\(400–700 nm)}                                  & {Intensity modulation and direct detection\\(e.g., OFDM, CAP)} & {\labelitemi\hspace{\dimexpr\labelsep+0.5\tabcolsep}Environmental monitoring\\\labelitemi\hspace{\dimexpr\labelsep+0.5\tabcolsep}Motion detection\\\labelitemi\hspace{\dimexpr\labelsep+0.5\tabcolsep}Spatial mapping}                   & {\labelitemi\hspace{\dimexpr\labelsep+0.5\tabcolsep}LoS-dependent\\\labelitemi\hspace{\dimexpr\labelsep+0.5\tabcolsep}Short range in NLoS\\\labelitemi\hspace{\dimexpr\labelsep+0.5\tabcolsep}Ambient light interference\\\labelitemi\hspace{\dimexpr\labelsep+0.5\tabcolsep}Issues with occlusion/diffusion} \\
FSO in ISAC              & {Infrared or Near-Infrared \\(740–1600 nm)}                     & {Laser-based detection\\(e.g., DCO-OFDM, LFM-CPM)}             & {\labelitemi\hspace{\dimexpr\labelsep+0.5\tabcolsep}Distance/angle sensing\\\labelitemi\hspace{\dimexpr\labelsep+0.5\tabcolsep}Velocity detection\\\labelitemi\hspace{\dimexpr\labelsep+0.5\tabcolsep}Atmospheric analysis}              & {\labelitemi\hspace{\dimexpr\labelsep+0.5\tabcolsep}Sensitive to turbulence\\\labelitemi\hspace{\dimexpr\labelsep+0.5\tabcolsep}Alignment precision required\\\labelitemi\hspace{\dimexpr\labelsep+0.5\tabcolsep}Complex maintenance for scalability}                                                         \\
Photonic Sensing in ISAC & {Optical signals \\(infrared/visible or localized wavelengths)} & Sharing the medium with sensing tools                          & {\labelitemi\hspace{\dimexpr\labelsep+0.5\tabcolsep}High-precision displacement\\\labelitemi\hspace{\dimexpr\labelsep+0.5\tabcolsep}Vibration detection\\\labelitemi\hspace{\dimexpr\labelsep+0.5\tabcolsep}Strain sensing (e.g., optical time-domain reflectometry)} & {\labelitemi\hspace{\dimexpr\labelsep+0.5\tabcolsep}High cost\\\labelitemi\hspace{\dimexpr\labelsep+0.5\tabcolsep}Challenges in scaling\\\labelitemi\hspace{\dimexpr\labelsep+0.5\tabcolsep}Sensitive to environmental noise\\\labelitemi\hspace{\dimexpr\labelsep+0.5\tabcolsep}Limited to static sensing}   
\end{tblr}

\end{table*}

\subsection{Hybrid RF-Optical ISAC Architectures}


Hybrid ISAC systems are typically realized through three architectural paradigms. The first is the modular loosely coupled design, where RF and optical chains operate in parallel under the coordination of a control or network layer. A representative example is the multi-modal platform in \cite{andrade2025demonstration}, which integrates VLC, FSO, and RF into a unified access network. The second is the tightly integrated physical-layer design, in which RF and optical front-ends are co-packaged with shared waveform generation and control interfaces. The third follows a functional division approach, where optical subsystems perform high-precision sensing or backhaul while RF links support broadband communication. A prime example is LiDAR-aided beamforming, where optical point clouds are processed to accelerate millimeter-wave beam selection and alignment \cite{zecchin2022lidar}. These paradigms differ in complexity and coupling depth but share the goal of exploiting complementary propagation characteristics and hardware advantages from both domains.




\subsubsection{Integration Architecture Models}

Efficient cross-modal task allocation is essential to fully exploit hybrid architectures. Joint resource allocation strategies aim to balance spectral efficiency with sensing accuracy. Recent studies have proposed joint power allocation methods for hybrid RF/VLC networks to optimize total system throughput under optical illumination constraints\cite{wu2021hybrid}. Furthermore, deep reinforcement learning (DRL) frameworks have been applied to dynamically adapt resource allocation between RF and optical links. For instance, a model-free DRL approach has been shown to surpass traditional optimization methods in hybrid WiFi/LiFi networks, achieving significant gains in achievable sum rate while reducing transmission power usage \cite{verma2025novel}.

\subsubsection{System-Level Performance Trade-offs}

Hybrid RF–optical systems must balance complementary strengths and weaknesses. Optical channels deliver high-resolution localization but require precise alignment and incur higher power consumption. In contrast, RF channels provide broader coverage and robustness to blockage, though at the cost of lower spatial resolution. Mode orchestration must also consider latency, mode-switching overhead, and redundancy for fail-safe operation. For example, while optical links can leverage NLoS reflections for environmental sensing as shown in \cite{zhang2024channel}, practical hybrid architectures often rely on RF links to maintain connectivity when optical paths are blocked, requiring intelligent context-aware coordination.


\subsubsection{Practical Implementation Challenges}

To effectively implement large-scale deployment of hybrid RF-optical ISAC architectures, several key engineering challenges must be addressed \cite{phuchortham2025survey}.

\paragraph{Inter-modal Synchronization} 

RF subsystems typically tolerate clock jitter on the order of sub-nanoseconds, whereas optical front-ends demand picosecond-level timing and phase coherence. Retroreflective optical OFDM experiments~\cite{cui2024retroreflective} demonstrated that mapping optical signals to the RF baseband can relax end-to-end timing alignment constraints in optical ISAC systems. In contrast, photonics-assisted W-band ISAC trials \cite{jia2024demonstration} showed that radar-aided beam training and alignment are required to maintain spatial co-alignment, resulting in additional control signaling that must be accounted for in practical system design. These observations indicate that practical hybrid transceivers should adopt dual-loop synchronization architectures that jointly stabilize the optical pulse train and the RF local oscillator.

\paragraph{Hardware Co-integration and Impairments} Co-packaging RF PAs, LNAs, optical modulators, phased-array elements, and beam-steering optics within a single transceiver imposes strict constraints on size, weight, power (SWaP), and thermal design. Studies on dual-hop mixed RF/FSO relaying systems have shown that soft-envelope limiter high-power amplifiers introduce a signal-to-noise-and-distortion ratio ceiling that cannot be mitigated by merely increasing optical power, leading to persistent outage floors under high peak-to-average power ratios \cite{8254526}. Consequently, co-packaged linearization techniques, such as digital pre-distortion, as well as photonic integrated circuits, are essential to mitigate such nonlinearities and maintain link performance.

\paragraph{Unified Channel Modeling} RF and optical channels exhibit fundamentally different propagation characteristics. RF links are typically modeled using small-scale fading distributions such as Rician or Nakagami, while optical channels suffer from turbulence-induced scintillation, beam misalignment, and weather-dependent attenuation. These disparities hinder the development of unified analytical frameworks for joint S\&C optimization. For vehicular scenarios, \cite{10937381} discussed ISAC-oriented channel conceptualization and modeling challenges, and emphasized the need to account for components such as LoS, specular reflections, and diffuse/clutter scattering when establishing a common modeling foundation.

\subsubsection{Summary and Discussion}

Hybrid RF–optical ISAC aims to combine the wide-area robustness and NLoS tolerance of RF with the high resolution and large bandwidth of optical systems, and recent progress in architectural integration, task allocation, and system optimization has demonstrated the potential of this paradigm. Nevertheless, several cross-domain challenges continue to constrain its scalability and efficiency. 

Waveform design remains a fundamental issue, as the two modalities exhibit inherently different signal constraints: optical channels are more susceptible to distortion and nonlinearities, necessitating tailored modulation formats and optimized power allocation \cite{10697102}, whereas RF channels demand robustness to fading and delay spread. Developing unified waveform strategies that simultaneously address coherence time, dispersion, and distortion profiles across both domains is still an open problem.

\begin{table*}
\centering
\caption{Summary of Waveform Design Methods with Orthogonal Resource Allocation in ISAC}
\begin{tblr}{
  width = \linewidth,
  colspec = {Q[70]Q[70]Q[210]Q[355]Q[180]Q[180]},
  row{1} = {c},
  cell{2}{1} = {c},
  cell{2}{2} = {r=3}{c},
  cell{2}{5} = {r=3}{},
  cell{2}{6} = {r=3}{},
  cell{3}{1} = {c},
  cell{4}{1} = {c},
  cell{5}{1} = {c},
  cell{5}{2} = {r=2}{c},
  cell{5}{5} = {r=2}{},
  cell{5}{6} = {r=2}{},
  cell{6}{1} = {c},
  cell{7}{1} = {c},
  cell{7}{2} = {r=2}{c},
  cell{7}{5} = {r=2}{},
  cell{7}{6} = {r=2}{},
  cell{8}{1} = {c},
  cell{9}{1} = {c},
  cell{9}{2} = {r=2}{c},
  cell{9}{5} = {r=2}{},
  cell{9}{6} = {r=2}{},
  cell{10}{1} = {c},
  vlines,
  hline{1-2,5,7,9,11} = {-}{},
  hline{3-4,6,8,10} = {1,3-4}{},
}
\textbf{Reference} & \textbf{Design Scheme} & \textbf{Specific Methods}                                                  & \textbf{Performance}                                                                          & \textbf{Advantages}    & \textbf{Disadvantages}         \\
\cite{9728752}              & Time division          & A mmWave enabled                                                           & {Target detection error reduced by 18.5\%, communication throughput 2.2 Gbps}             & {\labelitemi\hspace{\dimexpr\labelsep+0.5\tabcolsep}Simple implementation and control. 
\\\labelitemi\hspace{\dimexpr\labelsep+0.5\tabcolsep}Avoids interference between S\&C by isolating tasks in time}                                     & {\labelitemi\hspace{\dimexpr\labelsep+0.5\tabcolsep}Reduced temporal efficiency due to idle time slots\\\labelitemi\hspace{\dimexpr\labelsep+0.5\tabcolsep}Delayed operations for real-time tasks if time slots are too long}                    \\
\cite{9970330}    &                        & Integrated time-division JRC with federated edge learning                  & {Faster convergence and higher learning accuracy, especially in energy-limited scenarios} &    &                                                                       \\
\cite{10221868}         &                        & Cross-layer JRC scheduling policy optimization                             & {Trade-off between delay and power consumption while meeting detection accuracy}            &   &                                                 \\
\cite{10118989 }         & Frequency division     & {Optimize subcarrier and power allocation for S\&C} & {128 subcarriers, max power per subcarrier = 8W, significant power savings}                & {\labelitemi\hspace{\dimexpr\labelsep+0.5\tabcolsep}Avoids mutual interference by isolating tasks in frequency
\\\labelitemi\hspace{\dimexpr\labelsep+0.5\tabcolsep}Compatibility with existing FDMA-based systems}                                                & \labelitemi\hspace{\dimexpr\labelsep+0.5\tabcolsep}Inefficient spectral usage for systems with low resource demand in one domain                                                                                                                   \\
\cite{10570756}          &                        & Subcarrier and power allocation techniques                                 & {RMSE of range estimations improved at all CDR thresholds}                                &                                                                       &   \\
\cite{9424454}         & Spatial division       & {Pareto optimization framework for multi-antenna DFRC system}            & {Better performance under varied SINR constraints, converges in 6-7 iterations}           & {\labelitemi\hspace{\dimexpr\labelsep+0.5\tabcolsep}High spectral efficiency: Both S\&C coexist in the same frequency and time
\\\labelitemi\hspace{\dimexpr\labelsep+0.5\tabcolsep}Suitable for MIMO and mmWave systems with narrow beams} & {\labelitemi\hspace{\dimexpr\labelsep+0.5\tabcolsep}Complex hardware requirements (e.g., MIMO, beam steering)
\\\labelitemi\hspace{\dimexpr\labelsep+0.5\tabcolsep} Potential overlap or interference in dense deployment environments}          \\
\cite{9634053}        &                        & Closed-form solutions and iterative algorithms                             & {\labelitemi\hspace{\dimexpr\labelsep+0.5\tabcolsep}Algorithm 1: MI maximized, converges within about 10 iterations
\\\labelitemi\hspace{\dimexpr\labelsep+0.5\tabcolsep}Algorithm 2: Lower CRB}             &    &  \\
\cite{9359665 }          & Code division          & CD-OFDM with successive interference cancellation                          & {\labelitemi\hspace{\dimexpr\labelsep+0.5\tabcolsep}30.1 dB gain in the low SINR regime
\\\labelitemi\hspace{\dimexpr\labelsep+0.5\tabcolsep}A comparable sensing accuracy with a 3 dB SINR gain} & {\labelitemi\hspace{\dimexpr\labelsep+0.5\tabcolsep}High flexibility; both tasks can share time and frequency resources concurrently.\\\labelitemi\hspace{\dimexpr\labelsep+0.5\tabcolsep}Suitable for environments with multipath or clutter}                     & {\labelitemi\hspace{\dimexpr\labelsep+0.5\tabcolsep} Computationally intensive code design and decoding
\\\labelitemi\hspace{\dimexpr\labelsep+0.5\tabcolsep} Difficult to maintain strict orthogonality as system scales; risk of interference} \\
\cite{10003649}         &                        & {CPM codes for high-resolution radar images from airborne radar}         & Image quality enhanced, data rate up to 5 Mbit/s                                              &     &  
\end{tblr}
\label{tab:resource orthogonal}
\end{table*}

Task allocation likewise requires fine-grained adaptation to propagation conditions. Empirical studies indicate that assigning high-precision localization to VLC and control signaling to RF can improve both energy efficiency and robustness \cite{wu2021hybrid}, yet sustaining such gains in dynamic environments depends on real-time cross-layer coordination and predictive scheduling mechanisms capable of anticipating channel variation.

Equally critical are synchronization and physical-layer integration, where the disparity in timing requirements between RF and optical front-ends introduces additional latency. Experimental platforms, such as retroreflective optical OFDM \cite{cui2024retroreflective} and photonics-assisted W-band ISAC \cite{jia2024demonstration}, have shown that group-delay calibration and beam alignment impose non-negligible overhead. These findings highlight the need for dual-loop synchronization and SWaP-optimized co-packaged transceivers with adequate electromagnetic shielding.

Finally, the absence of a comprehensive channel modeling framework continues to limit system-level optimization, although the ISAC-V taxonomy \cite{10937381} provided an initial decomposition into LoS, target, specular-scattering, and diffuse-clutter components. 

\subsection{Lessons Learned}


The progression from RF to optical and then to hybrid ISAC highlights a broader shift toward spectrum-diverse, architecture-converged, and intelligence-driven wireless systems. Yet, the next stage of breakthroughs will likely rely not on isolated optimization within a single modality, but on their coordinated orchestration. This raises several key questions for the community:


\subsubsection{How Can We Build a Unified Co-Design Framework across Waveform, Hardware, and Protocol Layers}


Future architectures will need cross-domain performance metrics that simultaneously capture sensing resolution, communication throughput, and energy efficiency. Such a framework must establish shared optimization objectives, enabling systematic trade-offs between sensing fidelity and communication quality rather than treating them as independent targets.

\subsubsection{How Can Intelligence be Made Adaptive to Different Modalities}

Incorporating learning-based mechanisms to perform adaptive task offloading, resource allocation, and fault tolerance across RF and optical links is essential. RL and graph-based inference may enable context-aware reconfiguration under dynamic environmental and mobility conditions.

\subsubsection{How Can RF and Optical Systems be Synchronized and Calibrated in Real Time}
Emerging hardware platforms must achieve sub-nanosecond synchronization and joint calibration across RF and optical front-ends to overcome latency bottlenecks observed in experimental testbeds. This entails dual-loop timing architectures, coherent clock distribution, and integrated beam alignment procedures that minimize group-delay mismatches, ensuring stable operation in both static and mobile deployment scenarios \cite{cui2024retroreflective,jia2024demonstration}.

\section{Network Architecture from Single-Cell to Multi-Cell}\label{section3}

Early ISAC research predominantly focuses on single-cell architectures, where both S\&C functionalities are managed by a single BS. They leverage existing cellular and radar, providing a controlled environment for waveform design, interference management, and resource allocation. However, single-cell ISAC faces inherent limitations in coverage, spectral efficiency, and scalability due to the restricted sensing range and inefficient spectrum resource optimization.

The evolution toward multi-cell ISAC has been driven by the need for improved network robustness, sensing accuracy, and interference mitigation via collaborative sensing. Multi-cell ISAC enables joint S\&C among multiple nodes, resulting in novel challenges in waveform design, network topology, and cross-layer optimization. This section investigates the evolution from single-cell to multi-cell ISAC, highlighting key advancements in waveform design and network architectures.

\subsection{Single-Cell ISAC: A Shift in Waveform Design Methods}

In single-cell ISAC systems, the allocation of resources between S\&C introduces a fundamental coupling in system design. Sensing tasks, such as high-resolution radar imaging and precise target localization, require wide bandwidths and fine time–frequency resolution. However, allocating more resources to one function inherently constrains the other, a trade-off that is quantitatively captured by capacity–distortion theory and the CRB rate–region framework in Gaussian channels \cite{10147248}. The coexistence of these distinct requirements renders unified waveform design a central challenge.

\subsubsection{Waveform with Orthogonal Resource Allocation}


Early ISAC implementations often adopted orthogonal resource allocation across time, frequency, space, or code domains to decouple sensing echoes from communication signals. Such orthogonality simplifies transmitter design, receiver processing, and interference suppression, and has been widely studied for its integration and coordination gains \cite{9606831, liu2020joint}. Representative approaches are summarized in Table \ref{tab:resource orthogonal}. While orthogonal designs are effective for multi-user and multi-target separation, they typically limit spectrum reuse. More recent research has shifted toward non-orthogonal resource sharing, which concentrates energy more effectively and improves sensing SNR, but at the expense of increased transceiver complexity and tighter synchronization requirements.

\begin{figure*}[htp!]
    \centering
    \captionsetup{justification=raggedright}    
    \includegraphics[width=0.89\linewidth]{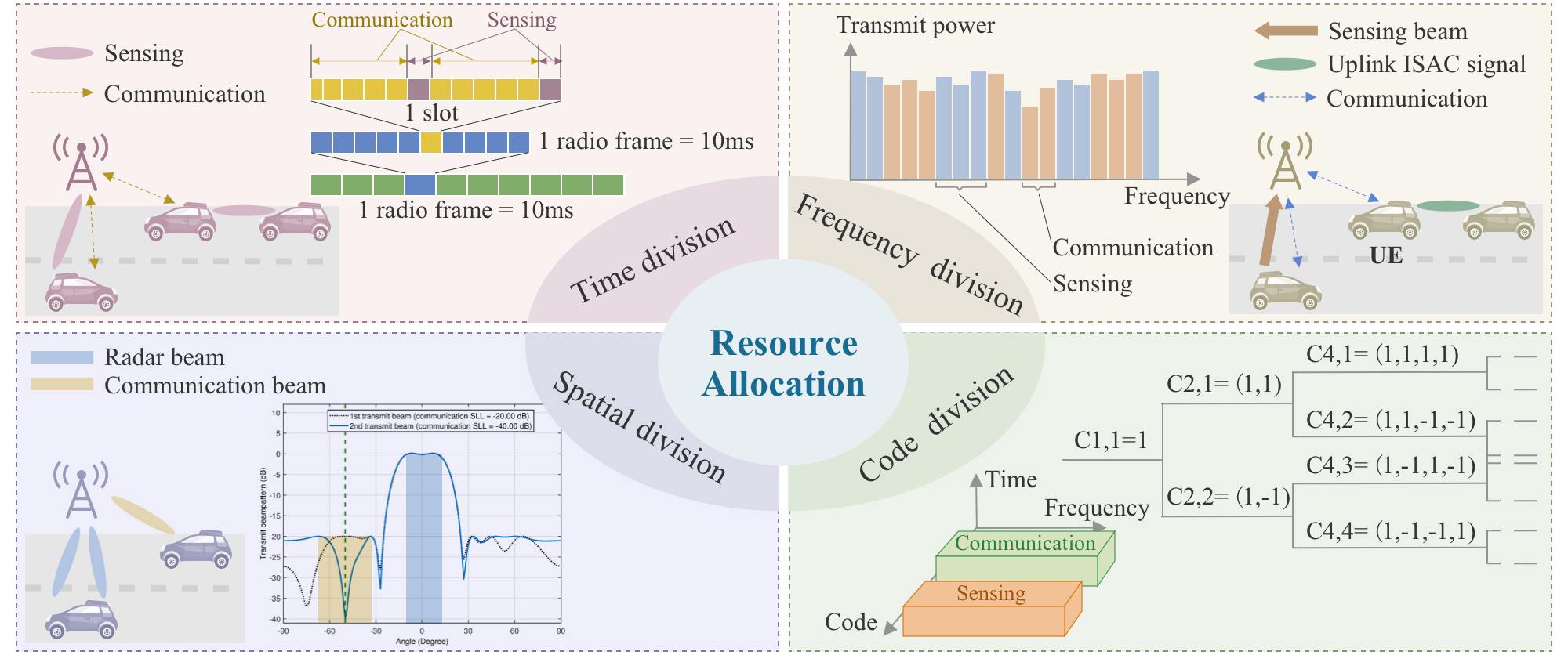}
    \caption{The figure illustrates orthogonal resource allocation in ISAC. In the time division section, a 10-ms radio frame allocates subframes for V2V/V2I communication and sensing. The frequency division section depicts uplink ISAC, where the UE transmits joint S\&C signals to the BS, optimized for sensing. In the spatial division section, ISAC beamforming with sidelobe control in a 10-antenna MIMO radar serves a UE at $-50^\circ$, maintaining a wide radar beam. Sidelobe levels fluctuate between $-40$ dB and $-20$ dB in the UE's direction. The code division section employs three-dimensional complementary coded scrambling to mitigate fading and interference, enhancing radar performance and data rates in multi-user environments.}
    \label{fig:4 divisions}
\end{figure*}

\paragraph{Time Division} The time division method involves allocating different time slots for S\&C waveforms, including pilot signals in communication systems and stepped-frequency pulses in radar systems. This approach is relatively easy to implement and can be integrated into existing systems with minimal modifications. Dynamically adjusting the resource allocation strategy based on the real-time requirements of S\&C is key to finding an optimal time-shared ISAC solution. For example, a mmWave-enabled cooperation algorithm was designed in \cite{9728752} based on a time division ISAC system for sharing raw sensing data among connected automated vehicles, as shown in Fig. \ref{fig:4 divisions}.


Recently, time division method has emerged as a leading strategy for ISAC deployment, with 5G implementations showcased in China, Korea, Europe, and the U.S. \cite{ZHU2023104262}. Major operators and device manufacturers have tested ISAC technology in commercial pilots. In 2023, ZTE conducted a successful 5G ISAC trial, using time-division multiplexing to transmit S\&C signals and detect UAVs as small as 0.01 m² within a 1 km range \cite{ZTE2024}. In October 2023, Huawei employed it at 4.9 GHz for ship tracking and micro-deformation sensing, achieving millimeter-level motion detection and sub-1 km/h speed errors at 20 km. In April 2024, China Mobile and ZTE deployed multi-station collaboration for continuous drone tracking, validating functionalities such as multi-target trajectory detection and persistent UAV monitoring \cite{ZTE20240412}. These deployments underscore the commercial readiness of time-division-based ISAC in complex real-world settings.

\paragraph{Frequency Division}
Benefiting from the large bandwidth of high-frequency bands, frequency division is deemed another solution for ISAC implementation.
It allocates different subcarriers for S\&C functions, distinguishing them in the spectrum domain. 
Fig. \ref{fig:4 divisions} shows a scenario where a user equipment (UE) sends uplink signals to a BS for joint S\&C \cite{10570756}. Non-overlapping subcarriers and optimized power enable precise sensing, supporting vehicle detection, and stationary identification near traffic lights.

Despite its advantages, frequency division still encounters several challenges. For instance, high-bandwidth radar systems often require multiple tightly packed subcarriers, which induce autocorrelation splitting and multiple spurious peaks, degrading sensing accuracy. 
Besides, cross-band integration is complicated due to parameter inconsistencies across different frequency ranges. To address these limitations, \cite{10129042} introduced a mutual-information-based framework for ISAC, leveraging signal MI as a unified metric for performance analysis. Such an approach enables more rigorous waveform design and balances the trade-offs between communication throughput and sensing quality.

\paragraph{Spatial Division}

Spatial division capitalizes on multi-antenna and massive MIMO techniques to separate S\&C signals in the spatial domain. Radar needs to project energy toward potential targets, whereas communication aims for optimal beamforming that delivers high SINR to specific users. These objectives are in conflict, as the strongest radar echo path may not coincide with the strongest communication channel.

A common approach is to control sidelobe levels, transmitting a main beam for radar while embedding communication bits into sidelobes. Though efficient, dynamic sidelobe fluctuations can degrade sensing accuracy in time-varying channels \cite{hassanien2015dual}. To address this conflict, an alternative solution is based on spatially extended orthogonal time-frequency space modulation\cite{9724198}. This framework discretizes the angle domain through a spatial spreading and despreading process, which effectively separates radar and communication waveforms. It streamlines the design of related estimation and detection problems while preventing the significant decline in communication performance that can occur when steering the beam toward the strongest radar signal path.

\begin{table*}
\centering
\caption{Constraints and Optimization Goals of Sensing-centric and Communication-centric Design}
\label{tab:waveform_constraints}
\begin{tblr}{
  width = \linewidth,
  colspec = {Q[58]Q[75]Q[345]Q[265]Q[65]Q[160]},
  row{1} = {c},
  cell{2}{1} = {c},
  cell{2}{2} = {c},
  cell{2}{5} = {c},
  cell{3}{2} = {c},
  cell{3}{5} = {c},
  cell{4}{2} = {c},
  cell{4}{5} = {c},
  cell{5}{2} = {c},
  cell{5}{5} = {c},
  cell{6}{1} = {r=4}{c},
  cell{6}{2} = {c},
  cell{6}{5} = {c},
  cell{7}{2} = {c},
  cell{7}{5} = {c},
  cell{8}{2} = {c},
  cell{8}{5} = {c},
  cell{9}{2} = {c},
  cell{9}{5} = {c},
  vlines,
  hline{1-2,6,10} = {-}{},
  hline{3-5,7-9} = {2-6}{},
}
\textbf{Design Scheme} & \textbf{Constraints}                             & \textbf{Mathematical Description} & \textbf{Description} & \textbf{Reference}             & \textbf{Application Scenario} \\
Sensing-centric        & Similarity & $| s-s_0 |^2 \leq \epsilon$  & Ensure waveform similarity to a reference with specified energy tolerance. & \cite{10121420}      & {\labelitemi\hspace{\dimexpr\labelsep+0.1\tabcolsep}Beamforming} \\
& PAPR & $\max(|s(t)|^2) / \mathbb{E}[|s(t)|^2]$& Limit amplitude range to improve amplifier efficiency and reduce distortion. & \cite{10695936}       & {\labelitemi\hspace{\dimexpr\labelsep+0.5\tabcolsep}UAV-Enabled Systems\\\labelitemi\hspace{\dimexpr\labelsep+0.5\tabcolsep}MIMO-OFDM Systems}       \\
& Frequency&$\int_{B_i} |S(f)|^2 df \leq P_{\max}$& Restrict waveform energy to specific bands for spectral coexistence.    & \cite{8804235 }               & {\labelitemi\hspace{\dimexpr\labelsep+0.5\tabcolsep}Subcarrier Assignment\\\labelitemi\hspace{\dimexpr\labelsep+0.5\tabcolsep}Spectrum Sharing}  \\
& Energy &$\int_{-\infty}^{\infty} |s(t)|^2 dt \leq E_{\max}$& Limit total transmitted energy based on system or detection range. & \cite{10445319 }              & {\labelitemi\hspace{\dimexpr\labelsep+0.5\tabcolsep}Beamforming\\\labelitemi\hspace{\dimexpr\labelsep+0.5\tabcolsep}Energy Efficiency} \\ 
{Communi\\cation-centric } & Fairness & 
{\(\max\limits_{F, P} \min\limits_{k \in K} \sum\limits_{n \in S} \ln \left(1 + \frac{|h_{n,k}|^2 f_{n,k} P_{n,k}}{u_k^2}\right)\)}
& Balance resources for consistent user performance under poor channel conditions. & \cite{9860724}                 & {\labelitemi\hspace{\dimexpr\labelsep+0.5\tabcolsep}V2X\\\labelitemi\hspace{\dimexpr\labelsep+0.5\tabcolsep}Distributed Systems for autonomous drones}                  \\
  & WMMSE & {\(\min\limits_{\mathbf{V}, \mathbf{B}, \mathbf{W}} \sum_{k} \alpha_k \left( \operatorname{Tr}(\mathbf{W}_k \mathbf{E}_k) - \log \det(\mathbf{W}_k) \right)\)} & Minimize weighted mean square error by optimizing beamforming for interference management. & \cite{10554642}              & {\labelitemi\hspace{\dimexpr\labelsep+0.5\tabcolsep}Intelligent
  Transportation Systems\\\labelitemi\hspace{\dimexpr\labelsep+0.5\tabcolsep}Multi-user System }\\
& SINR & 
\(\text{SINR} = \frac{\mathbf{w}^H \mathbf{H} \mathbf{s} \mathbf{s}^H \mathbf{H}^H \mathbf{w}}{\mathbf{w}^H (\mathbf{R}_{\text{intf}} + \sigma^2 \mathbf{I}) \mathbf{w}} \)

& Adjust beamforming to ensure robust signal detection under interference and noise. & \cite{8378754} & {\labelitemi\hspace{\dimexpr\labelsep+0.5\tabcolsep}V2X} \\
& Channel Capacity&
\(C = \log \det \left( \mathbf{I} + \frac{P}{N} \mathbf{H} \mathbf{H}^H \right)\)& Maximize data rate by optimizing power and noise levels. & \cite{xu2015radar} & {\labelitemi\hspace{\dimexpr\labelsep+0.5\tabcolsep}Smart City IoT\\} 
    \end{tblr}    
\captionsetup{justification=raggedright, singlelinecheck=false}
\caption*{$(*)$ $s_0$ is the reference waveform and $\epsilon$ controls waveform energy similarity. \( F \) and \( P \) represent the matrices for subcarrier allocation and power allocation, respectively. \( K \) is the set of communication users. \( S \) is the set of OFDM subcarriers. \(\mathbf{w}\) is the receive beamforming vector. \(\mathbf{H}\) represents the channel matrix. \(\mathbf{R}_{\text{intf}}\) is the interference covariance matrix. \(\sigma^2\) is the noise power.  \( \mathbf{I} \) is the identity matrix. \( P \) is the transmit power. \( N \) is the noise power.}
\end{table*}

\paragraph{Code Division}

In the code division mode, S\&C signals are carried by orthogonal/quasi-orthogonal sequences, occupying the same resources in the time-frequency domain to obtain higher resolution distance and speed estimation. A new code division scheme was proposed in \cite{10202575}. The scheme used three-dimensional complementary coded scrambling to combat wireless fading and mutual interference, acting as a soft isolator for integrated radar and communication systems in multi-user and multi-target scenarios, as shown in Fig. \ref{fig:4 divisions}.



In rapidly changing environments, balancing determinism and randomness in code division is particularly crucial. Deterministic waveforms can improve the ability to collect environmental information for a radar system, while communication systems rely on high randomness to achieve efficient data transmission and interference resistance. 
Therefore, careful planning of the codeword space is necessary for designing code division systems, ensuring that radar and communication functions can complement each other and reduce mutual interference while finding a balance between sensing performance and communication efficiency. 

\subsubsection {Unified Design}

 
As ISAC systems evolve, the inherent limitations of orthogonal designs, such as suboptimal spectrum utilization, restricted flexibility in resource reconfiguration, and the inability to fully exploit joint processing gains, have become increasingly evident. This motivates the transition toward unified waveform design, where S\&C share spectrum, hardware platforms, and processing chains under a joint resource allocation framework.


A unified waveform must balance conflicting requirements: communication waveforms are typically optimized for high spectral efficiency and data throughput, whereas sensing waveforms demand favorable ambiguity function properties and constant envelope characteristics to ensure accurate range, velocity, and angle estimation. The design problem is therefore formulated with task-specific constraints and multi-objective optimization goals, as summarized in Table \ref{tab:waveform_constraints}. Depending on the design priorities and waveform formats, three main strategies have emerged: sensing-centric, communication-centric, and joint co-design approaches.

\paragraph{Sensing-Centric Design}

In sensing-centric designs, communication functionality is embedded into a radar-optimized waveform, with a primary focus on preserving high-fidelity sensing. 
A representative waveform is the chirp-based frequency-modulated continuous wave (FMCW) signal, denoted by
\begin{equation}
s_{\text{FMCW}}(t) = \exp\left[j 2\pi \left(f_0 t + \frac{k}{2} t^2\right)\right],
\end{equation}
where $f_0$ is the central frequency and $k$ denotes the chirp rate. The linear frequency sweep provides fine range resolution and robustness to noise. Compared with OFDM, FMCW generally requires much lower sampling rates, which simplifies receiver processing. However, this advantage comes at the expense of significantly reduced data transmission rates, making FMCW more suitable for sensing-centric designs \cite{sturm2011waveform}.

Additionally, index modulation schemes allow information to be conveyed through antenna or frequency selection without altering the primary radar waveform. Recent studies have explored these embedding and modulation techniques, emphasizing trade-offs in symbol rate and beam fidelity.  For example, \cite{9345999} demonstrated a spatial modulation-based ISAC prototype achieving BER on the order of $10^{-2}$ at 15 dB SNR while enhancing angular resolution through spatial agility. Similarly, \cite{liu2020joint} proposed a joint beamforming framework for multi-user MIMO ISAC, maintaining a radar beam pattern that closely matches the optimal radar-only design while achieving near-optimal communication sum rate. However, sensing-centric designs generally offer limited communication capacity and impose higher receiver processing complexity.

\paragraph{Communication-Centric Design}

Conversely, communication-centric designs superimpose sensing functions onto existing communication waveforms, most notably OFDM and orthogonal time–frequency space (OTFS) \cite{8288677} modulation \cite{wei2021orthogonal,10638525}.

OFDM, widely adopted in 5G NR and IEEE 802.11 standards, provides high spectral efficiency and robustness against multipath fading by transmitting data over multiple orthogonal subcarriers. This facilitates flexible allocation of time–frequency resources for simultaneous sensing. The baseband OFDM signal can be expressed as

\begin{equation}      
    \begin{split}
   \!\!\!\! s_{\text{OFDM}}(t)\! =\! \sum_{m=1}^{N_s}\sum_{n=1}^{N_c} d_{m,n}\, e^{j2\pi n\Delta f\, t}\, \text{rect}\!\left(\!\frac{t-(m-1)T}{T}\right),
    \end{split}
\end{equation}
where \(N_s\) is the number of OFDM symbols, \(N_c\) is the number of subcarriers, \(d_{m,n}\) are the data symbols, \(\Delta f\) is the subcarrier spacing, \(T\) is the OFDM symbol period, and \(\text{rect}(\cdot)\) is the rectangular function.


For radar processing, the range-Doppler matrix is constructed  to achieve effective separation and joint estimation of delay and Doppler as expressed by
\begin{equation}      
\!\!Y(m,n)\!=\!\sum_{l=1}^{L} \alpha_l\, d_{m,n}\, e^{-j2\pi n\Delta f\, \tau_l}\, e^{j2\pi mT\, \nu_l} + W(m,n),
\end{equation}
where \(\tau_l\) and \(\nu_l\) represent the delay and Doppler shift, respectively, \(\alpha_l\) is the reflection coefficient for the \(l\)-th target, \(W(m,n)\) is the frequency-domain noise, and \(m\) and \(n\) are the symbol and subcarrier indices, respectively. 
Applying two-dimensional (2D) processing (e.g., a 2D FFT) to \(Y(m,n)\) enables joint estimation of target range and velocity. Despite these advantages, OFDM suffers from high sidelobe levels and a poor ambiguity function, which limits its sensing accuracy. To mitigate this, \cite{10255745} optimized subcarrier power allocation based on MI, enabling an effective trade-off between sensing and communication performance. Symbol-level precoding was introduced in \cite{10771629} to significantly suppress range–Doppler sidelobes (e.g., achieving tens of dB integrated sidelobe level reduction in simulations), thereby enhancing target detection. Null-space projection further improves sensing performance without compromising communication quality \cite{10661215}. Nevertheless, the sensitivity of OFDM to Doppler shifts under high-mobility conditions remains a major limitation.

To address this, OTFS has emerged as a compelling alternative for ISAC applications \cite{10638525}. The OTFS signal can be expressed as
\begin{equation}      
    \begin{split}
    s_{\rm OTFS}(t) = \int \int s(\tau, \nu) \cdot \phi(t - \tau) \cdot e^{j2\pi\nu t} \, d\tau \, d\nu,
    \end{split}
\end{equation}   
where $s(\tau, \nu)$ is the 2D modulation symbol at time-shift $\tau$ and frequency-shift $\nu$, $\phi(t - \tau) $ denotes the pulse shape function, $e^{j2\pi\nu t} $ is the complex exponential carrier frequency modulation term. By inherently mapping information symbols to delay–Doppler coordinates, OTFS couples communication signaling with parameter estimation, resulting in a distinctive “pushpin-shaped” ambiguity function that enables high resolution in both delay and Doppler.


Recent advances have further validated the dual-functional superiority of OTFS. Multi-objective optimization has been applied to balance the trade-off between subcarrier adaptability and spectral efficiency \cite{10257601}. To enhance sensing performance, OTFS waveform designs with sidelobe control have been proposed to substantially suppress delay--Doppler sidelobes in simulations, achieving sidelobe levels as low as -35 dB while maintaining reliable communication \cite{10521551}. By reusing downlink echoes for uplink channel estimation, OTFS further demonstrated robust operation without requiring dedicated sensing waveforms \cite{10941992}. Moreover, delay--Doppler structural priors were leveraged to refine grid resolution and reduce pilot overhead for channel estimation~\cite{10978674}. A unified iterative framework was later developed to jointly perform parameter association, channel estimation, and signal/data detection, which consistently improved estimation and detection performance in high-mobility environments~\cite{10409528}. These contributions collectively demonstrate that OTFS offers a promising foundation for tightly integrating S\&C in dynamic wireless systems.

\paragraph{Joint Design}


In contrast to separate optimization strategies, joint S\&C design formulates the objectives of the two functions within a unified mathematical framework. This formulation enables a direct characterization of their coupled behaviors and provides a principled view of how sensing and communication interact over shared wireless resources.

The motivation for unified formulations arises from the fundamental mismatch between the waveform requirements of the two functionalities. Sensing typically benefits from deterministic and energy-concentrated signals that enhance estimation precision\cite{he2012waveform}, whereas communication favors random and spectrally efficient signaling to maximize data throughput and robustness\cite{cover1999elements}. If optimized independently, these different requirements often yield incompatible designs, thereby motivating joint objectives that explicitly encode the shared resource constraints and physical coupling\cite{9737357,10418473,9585492}.

Among various unified metrics, MI is widely adopted because it simultaneously characterizes the recoverable information in communication and the observability of target-induced perturbations required for sensing. MI is also grounded in information-theoretic limits, making it resilient to practical impairments such as partial channel knowledge, interference, and clutter—conditions frequently encountered in ISAC deployments\cite{10466681}.

The joint optimization problem is typically formulated as a weighted MI maximization 
\begin{equation}
\begin{aligned}
    \max & \quad \rho I(X; Y_c | H_c) + (1 - \rho) I(H_s; Y_s | X) \quad  ,
\end{aligned}
\end{equation}

\noindent where \(X\) denotes the jointly designed transmit signal, \(\rho \in [0, 1] \) is a weighting factor, I denotes the MI, and \(Y_c\), \(Y_s\), \(H_c\),  and \(H_s\) represent the received signals and channel/target impulse responses for communication and sensing, respectively. This formulation identifies the Pareto-optimal MI frontier, illustrating the fundamental trade-offs between S\&C.

Given that power and bandwidth are consumed by sensing and communication, ISAC systems inherently require a unified resource allocation framework. Sensing accuracy improves with transmit energy and bandwidth, whereas communication simultaneously requires rate, latency, and reliability guarantees. Because these demands originate from the physical coupling of a dual-functional transceiver, resource allocation emerges as a central design challenge.

\begin{equation}
\begin{array}{l}
\!\!\!\displaystyle \max_{\mathbf{p},\mathbf{b}} \quad \text{Sensing QoS (detection, localization, tracking)} \!\\
~\text{s.t.} \quad R_{\mathbf{p},\mathbf{b}} \ge R,\quad \mathbf{1}_M^T \mathbf{p} = P_{\text{total}},\quad \mathbf{1}_M^T \mathbf{b} = B_{\text{total}},
\end{array}
\end{equation}
\noindent where \(R_{\mathbf{p},\mathbf{b}}\) denotes the achievable communication sum-rate, required to exceed a minimum threshold \(R\). The vectors \(\mathbf{p}\) and \(\mathbf{b}\) represent the power and bandwidth allocated across \(M\) sensing targets or communication users, subject to total resource constraints \(P_{\text{total}}\) and \(B_{\text{total}}\), respectively.

The sensing objectives originate from classical radar detection and estimation theory \cite{van2002optimum}, forming the fundamental limits for detecting, localizing, and tracking objects under noisy reflections. Hence, these criteria appear consistently in ISAC optimization.

\begin{itemize}

\item Detection Quality of Service (QoS):

\begin{align*}
       P_D = 1 - F_{\chi^2_2}\left(\frac{2\delta}{N_r\,\sigma_r^2\,\left(1 + p_q\,\epsilon_q\right)}\right),
\end{align*}

where $P_D$ denotes the probability of correctly declaring the presence of a target; $\epsilon_q$ is the normalized sensing channel gain associated with transmit power $p_q$ and array gain; $\delta$ is the detection threshold; $N_r$ is the number of receive antennas; and $\sigma_r^2$ is the noise variance. Physically, $P_D$ measures the system’s ability to detect a target under noise and clutter. A higher $P_D$ directly reduces miss detection in safety-critical applications (e.g., collision avoidance or intrusion monitoring), which explains why power allocation in low-SNR regimes often prioritizes reliable detection.

\item Localization QoS:
\[
\text{CRB}(d_q) = \frac{\beta_1}{p_q\,|c_q|^2\,b_q}, \qquad \text{CRB}(\theta_q) = \frac{\beta_2}{p_q\,|c_q|^2}
\]
\[
\rho_q = \frac{\omega_r}{\text{CRB}(d_q)} + \frac{\omega_\theta}{\text{CRB}(\theta_q)},
\]

The CRB specifies the minimum achievable variance of unbiased estimators for range $d_q$ and angle $\theta_q$. These bounds decrease with transmit power $p_q$, in the case of range estimation, with allocated bandwidth $b_q$. The constants $\beta_1$ and $\beta_2$ depend on waveform parameters and array geometry. A larger bandwidth $b_q$ improves delay (range) resolution, whereas a higher transmit power $p_q$ enhances angular resolution through improved SNR. The composite metric $\rho_q$ thus reflects the achievable localization accuracy of an ISAC transceiver and is essential for beam alignment, user positioning, and environment reconstruction.

\item Tracking QoS:
\[
\text{PCRB}_{\xi_{q,n}} = \mathbf{J}_{\xi_{q,n}}^{-1}(p_{q,n},\,b_{q,n}),~
\rho_{q,n} = \text{trace}\left(\text{PCRB}_{\xi_{q,n}}\right)
\]
where $\xi_{q,n}$ is the dynamic target state (e.g., position and velocity), and $\mathbf{J}_{\xi_{q,n}}$ is the corresponding Fisher information matrix. The posterior CRB (PCRB) characterizes the tightest uncertainty bound for estimating time-varying trajectories given the sensing update rate. A smaller PCRB indicates more stable and accurate tracking, which is particularly critical in high-mobility environments. Since ISAC systems must simultaneously accommodate per-slot communication constraints (e.g., latency and throughput), the transmit power $p_{q,n}$ and bandwidth $b_{q,n}$ available for sensing directly determine the achievable tracking accuracy and update rate.

\end{itemize}

Recent studies have introduced frameworks addressing joint optimization challenges. Closed-form characterizations of ISAC metrics such as false-alarm and detection probabilities under power allocation were provided in \cite{10124135}, offering insights into how transmit energy affects sensing accuracy. A flexible resource allocation strategy for jointly distributing power and bandwidth was developed in \cite{10387517}. Moreover, an optimal sum-of-ratios-based solution was proposed in \cite{10217150} to coordinate power, time, and quantization budgets across distributed devices. These developments demonstrate that ISAC optimization objectives arise naturally from physical constraints, estimation limits, and shared-resource coupling, rather than being imposed by arbitrary design preferences.

Practical deployment considerations further underscore the relevance of these formulations. Hardware impairments, synchronization offsets, finite-rate control signaling, and dynamic traffic loads introduce deviations from idealized assumptions, affecting both sensing and communication. Because these non-idealities influence both functionalities simultaneously, joint S\&C optimization aligns closely with the practical behavior of integrated transceivers\cite{10418473,10273396}.

Overall, the existing studies indicate the increasing importance of developing efficient and adaptive joint resource allocation mechanisms in ISAC, which provide a foundation for scaling ISAC capabilities toward larger and more heterogeneous deployments.


\subsubsection {Summary and Discussion}


The initial phase of ISAC evolution is primarily characterized by the transition from orthogonal to partially integrated resource allocation, where S\&C functions share spectrum resources within controlled interference levels. The feasibility of dual-functional waveforms was experimentally validated in \cite{9903001}, where real-time prototypes confirmed that practical trade-offs between the two domains could be achieved in controlled test environments. These early studies demonstrated that simultaneous transmission and sensing were technically feasible, although the coexistence of the other inevitably constrained the performance of each domain.

 \begin{figure*}[!htp]
        \centering
        \includegraphics[width=0.9\linewidth]{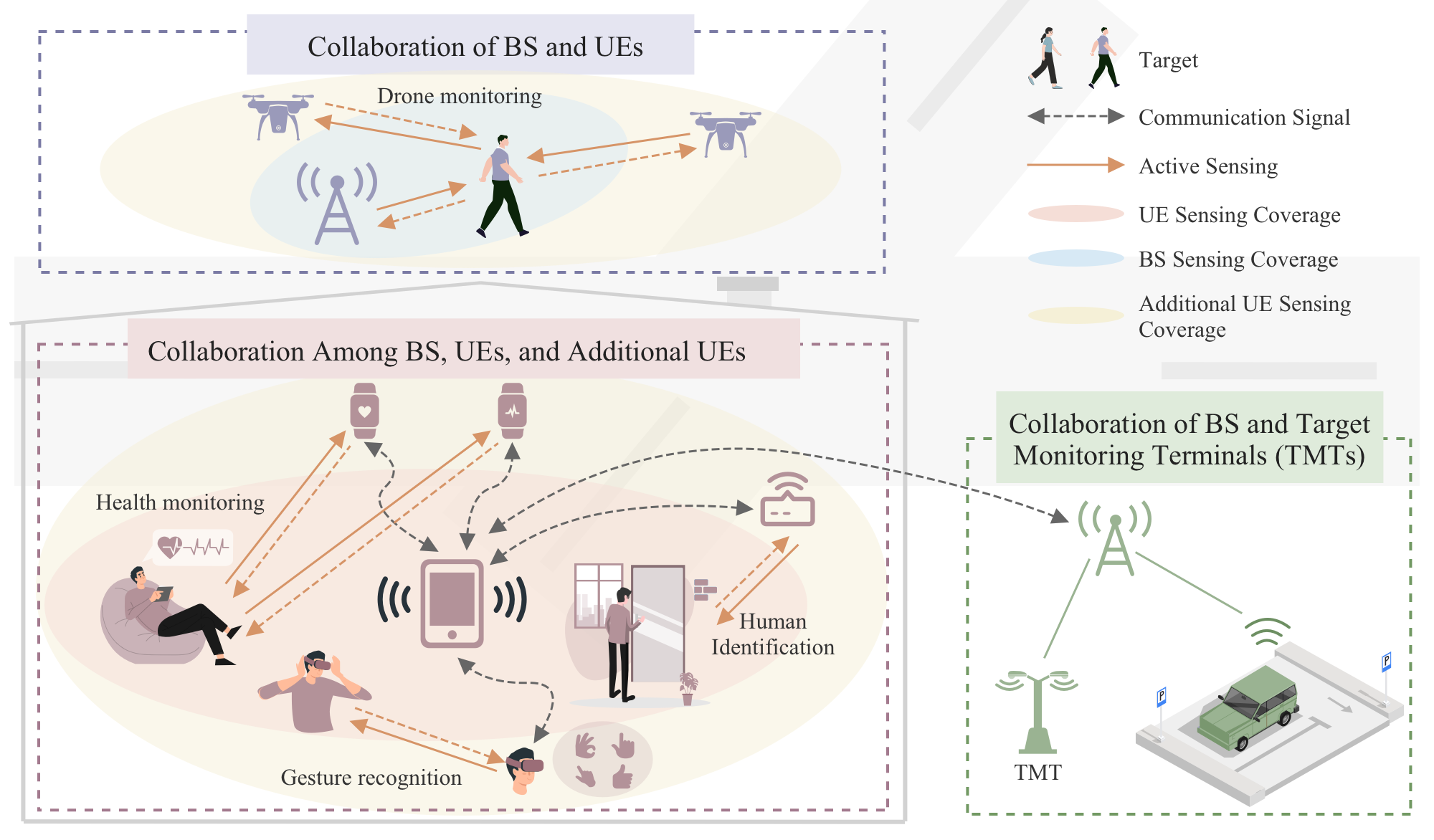}
        \captionsetup{justification=raggedright} 
        \caption{Topologies of single-cell networks, including collaboration between BS and UEs, and BS collaboration with target monitoring terminals.}
        \label{fig:Topologies of Single-Cell Network}
    \end{figure*}  

Subsequent research efforts shifted towards developing waveform designs capable of leveraging common hardware architectures and unified signal processing algorithms\cite{10418473}. This stage emphasized the need for a holistic framework that not only maximizes individual link performance but also coordinates cross-domain resource utilization for global system efficiency. Nevertheless, fundamental challenges persisted, particularly in characterizing performance boundaries and in modeling the complex mechanisms of cross-domain interference. While studies such as \cite{liu2023fundamental} made initial progress in analyzing interference in shared-spectrum ISAC systems, this area remained underdeveloped compared to traditional communication or radar theory.

From a practical standpoint, reusing communication waveforms (e.g., OFDM-based 5G NR signals) offers clear advantages in terms of compatibility and cost efficiency\cite{8288677}. However, sensing performance remains suboptimal when relying on communication-centric designs. This mismatch underscores the necessity of developing task-driven and sensing-optimal ISAC strategies, especially under imperfect synchronization, dynamic channels, and heterogeneous task demands.


In summary, the evolution of ISAC waveform design has progressed from feasibility demonstrations to integrated optimization, establishing a foundation for practical dual-functional operation. However, key challenges remain unresolved. Interference characterization in shared-spectrum environments is still not fully understood, performance limits have yet to be rigorously quantified, and the design of sensing-optimal waveforms under realistic constraints requires further exploration. Addressing these issues is essential for moving beyond ad hoc reuse of communication signals toward systematically engineered ISAC frameworks capable of supporting diverse 6G applications.

\subsection{Topology Evolution from Single-Cell to Multi-Cell Network}

Having established the fundamental waveform design principles within single-cell ISAC systems, our discussion was extended to multi-cell architectures, where cooperation among distributed nodes introduced additional opportunities and complexities.


\subsubsection{Types of Sensing Entities}

\paragraph{BS} In ISAC networks, BSs (including macro BSs and micro BSs) provide connections between UEs and the core network. They manage wireless access, including sending and receiving S\&C signals as well as controlling and managing UEs.

\paragraph{UE} UEs are equipped with specialized sensors to gather data about the environment or surrounding objects and communicate with BSs using wireless signals. Owing to mobility and flexibility, multiple UEs collaboratively extended sensing coverage. For example, satellites in the global navigation satellite system (GNSS) and BSs in current cellular networks can utilize device-based sensing technology to locate mobile devices \cite{liu2022networked}.

\paragraph{Additional UE} 
Additional UEs correspond to those terminal devices that cannot connect to the BS directly, including smartwatches and fitness trackers. Equipped with cameras, inertial, and environmental sensors, they connected to smartphones or other UEs via Bluetooth/Wi-Fi and enhanced area awareness and mobility coverage at low cost and with high portability.

\paragraph{Summary and Discussion}

Multiple sensing entities cooperated to rapidly fuse multi-perspective observations. The aggregated data were delivered over backhaul to processing nodes for inference, improving detection accuracy, parameter estimation, and effective coverage. Representative topologies included single-cell, multi-cell, and space–air–ground integrated network, as detailed below.


\subsubsection{Topology of Single-Cell Network}

Single-cell topology enables rapid integration of environmental information with minimal modifications to existing infrastructure. Standalone BSs provided direct services to terminals without reliance on the legacy core network. As illustrated in Fig. \ref{fig:Topologies of Single-Cell Network}, the BS acts
as the master node, while UEs and additional UEs act as subordinate nodes for cooperative sensing within the coverage area. Full-duplex operation is assumed to support simultaneous transmission and reception, with sensing integrated to optimize resource use and network performance.


\paragraph{Collaboration between BS and UEs}

The BS and UEs collaboratively fuse signals to enhance sensing. In downlink sensing, the BS may combine uplink probes from UEs with locally received echoes, or fuse downlink echoes with UE-side reflections. Since BS and UE are spatially separated, relative delays and Doppler shifts arise, which require estimation. Triangulation-based techniques are often applied to resolve ambiguities. Additional UEs equipped with sensors can measure vital signs or motion state. They forward processed features or raw measurements to peer UEs and the BS through sidelink reference signals. This cooperation extends the sensing footprint and enables more flexible coverage.




 \begin{figure*}[!htp]
    \centering
    \captionsetup{justification=raggedright} 
    \includegraphics[width=0.90\linewidth]{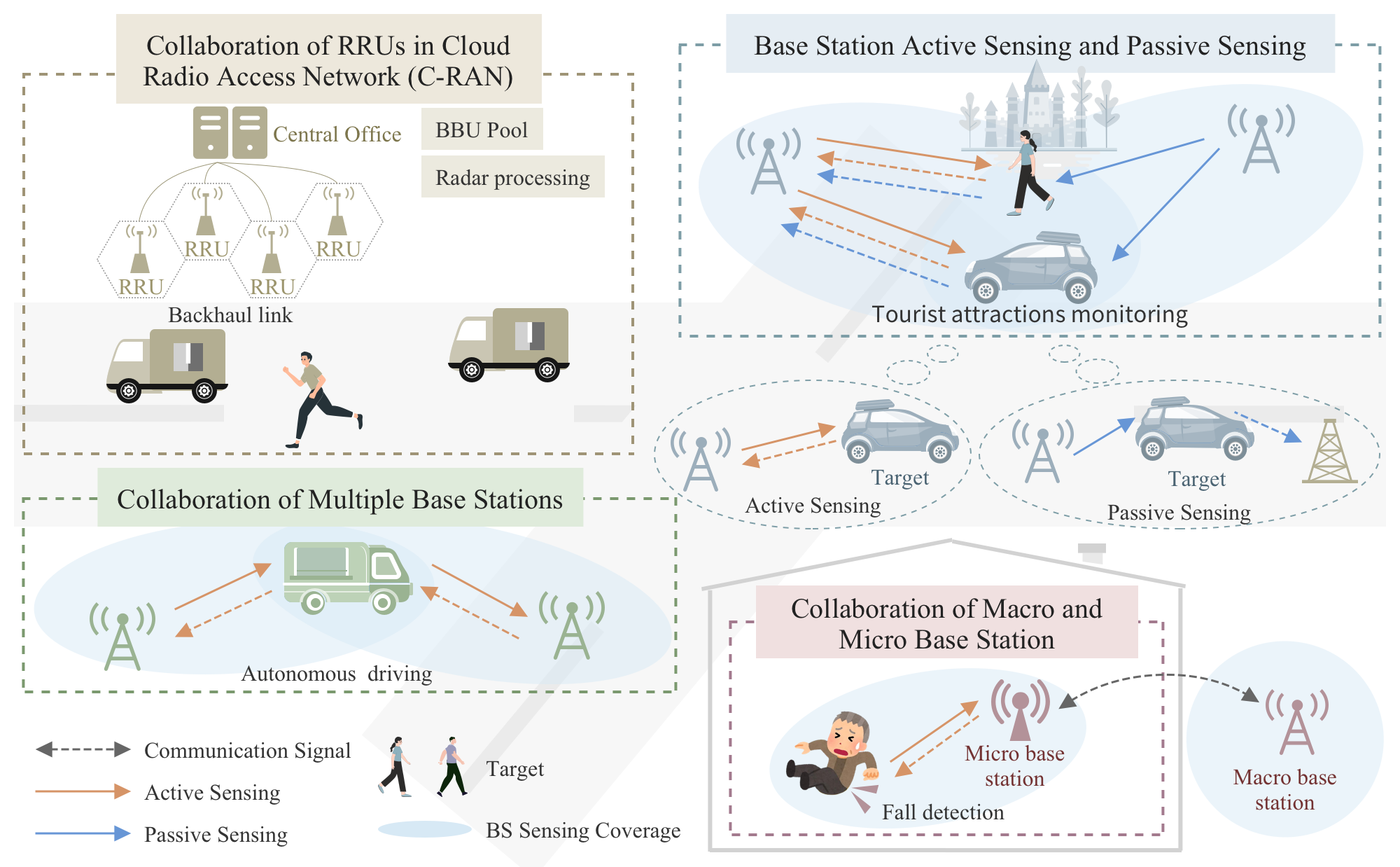}
    \caption{Topologies of multi-cell Network, illustrating: collaboration of RRUs in C-RAN, coordination among multiple BSs for joint monostatic and bistatic sensing, integration of monostatic and bistatic sensing at a single BS, and cooperation between macro and micro BSs.}
    \label{fig:Topologies of Multi-Cell Network}
\end{figure*}

\paragraph{Collaboration of BS and Target Monitoring Terminals}
Target monitoring terminals are passive sensing nodes (e.g., passive radar) without active transmission. Deployed across target areas and connected to BSs over low-latency links, they expand coverage with minimal interference.

Single-cell ISAC systems face inherent challenges. Sensing accuracy and coverage are limited to the BS’s service area, especially in dense urban scenarios. The coexistence of S\&C introduces intra-cell interference, such as uplink/downlink and UE-to-UE interference, degrading signal separability and efficiency. Moreover, scalability is constrained, as a single BS may become overloaded when the sensing requirement grows.


These limitations highlight the need for tailored strategies such as interference mitigation, coordinated beamforming, and asynchronous sensing support. A key issue is intra-cell interference arising from overlapping radar and communication functionalities at centralized BSs. In \cite{han2024cellular}, a cellular network–based multistatic ISAC configuration was investigated, where cooperative sensing across multiple base stations effectively reduced interference power by over 10~dB compared to monostatic benchmarks, while maintaining reliable sensing performance. To further improve operation, \cite{10577579} proposed a coordinated multipoint framework incorporating distributed beamforming and joint SINR optimization, which achieved communication SINR gains of up to 8 dB without degrading radar detection accuracy. Beyond waveform-level coordination,  \cite{9349171} studied uplink sensing under asynchronous transceivers and constrained antenna resources. Cross-antenna cross-correlation accurately estimates delays and Doppler shifts without strict synchronization, effectively resolving multi-target ambiguities by exploiting spatial correlation across receive antennas.


Overall, the transition from single-cell to multi-cell ISAC systems reflected a progression from localized integration toward large-scale collaborative architectures. While BSs acted as central nodes and UE devices extended coverage and functionality, persistent interference, scalability, and coverage limitations motivated multi-cell coordination.

\subsubsection{Topology of Multi-Cell Network}



Multi-cell ISAC architectures extend sensing coverage and enhance spatial diversity through coordination among multiple BSs (or gNBs in 5G NR terminology), as shown in Fig. \ref{fig:Topologies of Multi-Cell Network}. Overlapping service areas enable multi-perspective data fusion, thereby enhancing robustness and granularity.

\begin{figure*}[!htp]
        \centering
        \includegraphics[width=0.95\linewidth]{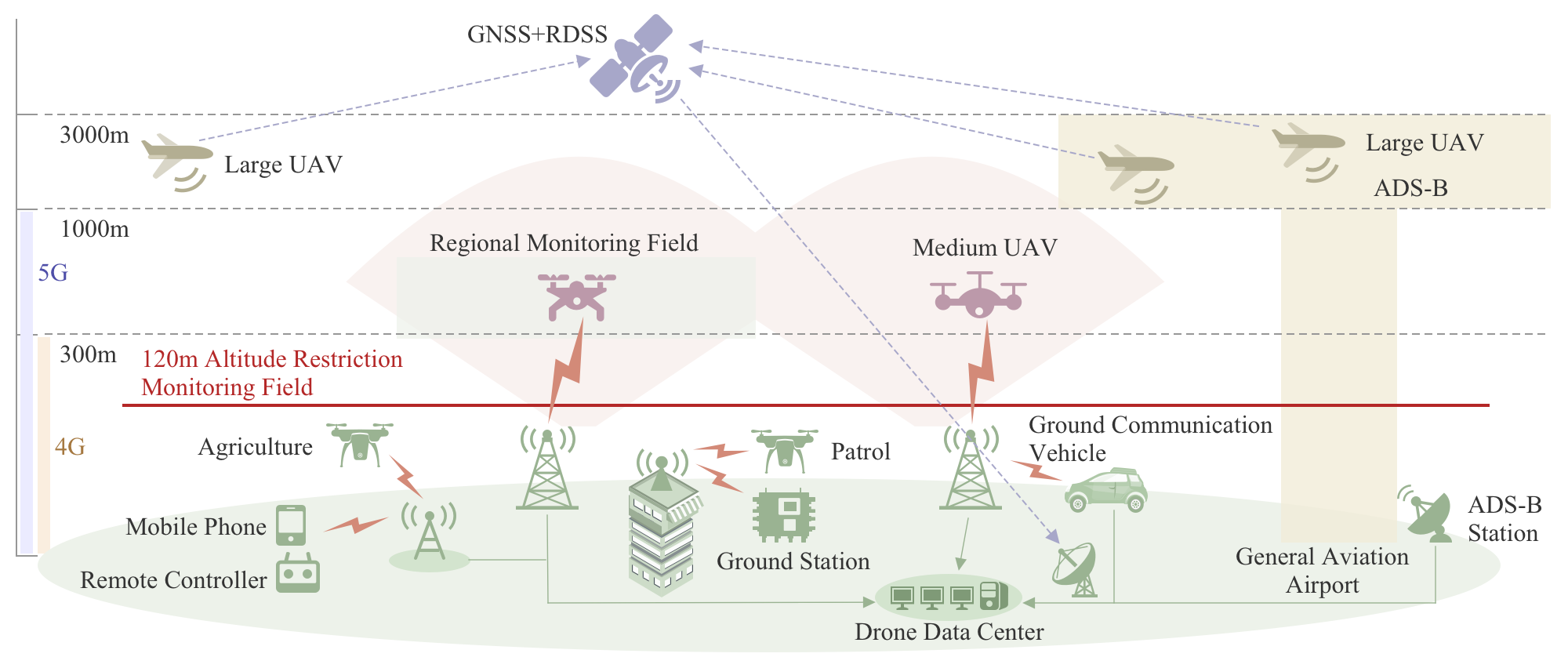}
        \caption{Topologies of Space-air-ground Integrated Network}
        \label{fig:SAGIN}
\end{figure*}

\paragraph{Collaboration of Multiple BSs}


In cooperative ISAC networks, multiple BSs are jointly operated in both monostatic and bistatic modes. Monostatic sensing relies on locally received echoes, whereas bistatic sensing exploits echoes from adjacent BSs or UEs, thereby forming multi-static configurations that enhance angular resolution and robustness against occlusion. To address inter-cell interference in dense deployments, coordinated beamforming and interference nulling have been proposed. For instance, \cite{10769538} developed a stochastic geometry-based framework showing that cooperative ISAC significantly enhances localization accuracy and communication rates (e.g., achieving up to 78\% rate gain in simulations) compared to non-cooperative baselines. Beyond interference management, clustering strategies play a critical role. Building upon these insights, \cite{10969840} proposed a data-aided bistatic sensing framework, which utilizes decoded communication data symbols to refine channel and target parameter estimation, thereby improving spectral efficiency without requiring excessive pilot overhead.



\paragraph{Collaboration of Macro and Micro BSs} 

Heterogeneous collaboration between macro and micro BSs is particularly effective in urban and indoor scenarios, where macro BSs suffer from penetration loss and coarse coverage granularity. Strategically deployed micro BSs (e.g., femtocells) provide localized perception and support seamless indoor–outdoor integration. A hybrid-access framework in \cite{10901518} incorporating social trust and operator collaboration improved offloading efficiency and energy efficiency in simulations (e.g., offloading efficiency increased from 19.59\% to 45.16\% under a representative setting, while energy consumption rating decreased from 32.37 to 22.89 W/Mbps). Moreover, macro BSs can serve as MEC fusion centers to aggregate sensing data from nearby micro BSs for joint processing and beamforming optimization. Coordinated multi-BS sensing, as discussed in \cite{10273396}, has been shown to significantly improve detection accuracy and extend sensing range in dynamic environments.

\paragraph{Collaboration of RRUs in Cloud Radio Access Network (C-RAN)}


C-RAN, introduced around 2009, further advances the multi-cell paradigm by decoupling and centralizing BS functionalities. In LTE and 5G NR, a BS (gNB) can be functionally split into centralized units for higher-layer baseband processing, distributed units for lower-layer real-time tasks, and remote radio units (RRUs) for analog transmission and reception  \cite{3gpp_tr38801, TS38401}. To address full-duplex challenges, some RRUs are designated as dedicated receivers. When combined with massive MIMO and cooperative beamforming, C-RAN enables large-scale, coordinated wide-area sensing. Joint signal- and data-level fusion in C-RAN has been shown to improve both communication throughput and sensing accuracy \cite{9737357}. In high-density networks, \cite{10735119} investigated optimizing cluster size, user density, and spatial resource allocation to fully leverage the performance gains.

\paragraph{Summary and Discussion}

The progression from distributed BS cooperation to heterogeneous macro–micro deployments and centralized C-RAN frameworks highlights the growing complexity of multi-cell ISAC. While these architectures offer extended coverage, spatial diversity, and improved sensing accuracy, they also introduce challenges in inter-cell synchronization, interference management, and coordination overhead. Advanced signal processing methods, such as compressed sensing and background subtraction, have been applied to recover perception parameters and mitigate multipath distortion \cite{rahman2019framework}. A stochastic geometry model in \cite{10735119} demonstrated that inter-cell coordination (specifically interference nulling) effectively enhanced both the average data rate and sensing performance by suppressing cross-link interference. Moreover, joint beamforming and SINR optimization were proposed in \cite{10304081} to maximize the worst-case radar SNR, thereby achieving seamless sensing coverage under strict communication constraints. These findings underscore the need for unified cross-layer frameworks that can balance S\&C demands in dynamic multi-cell environments.

\subsubsection{Space-Air-Ground Integrated Network}

The space–air–ground integrated network has been proposed to provide seamless connectivity and joint S\&C services across multiple layers. As shown in Fig. \ref{fig:SAGIN}, this architecture supports applications such as environmental monitoring, disaster response, and remote communications in areas where conventional infrastructure is limited. By integrating satellites, aerial platforms, and terrestrial networks, it not only extends coverage but also enables large-scale, real-time data acquisition and situational awareness across diverse geographic environments\cite{yuan2025groundskyarchitecturesapplications}. 

\paragraph{Collaboration of Space and Ground} 
In space–ground collaboration, satellites in low Earth orbit (LEO), medium Earth orbit (MEO), and geostationary orbit (GEO) communicate with ground networks and stations. This collaboration enhances global situational awareness and enables sensing technologies covering entire continents and oceans. LEO satellites, located at altitudes of 160-2,000 km, provide low-latency connectivity and enable ultra-dense satellite constellations, thereby facilitating real-time Earth observation and high-frequency environmental monitoring. MEO satellites balance coverage and latency, serving global navigation, IoT connectivity, and wide-area communication. Meanwhile, GEO satellites, positioned at approximately 35,786 km, provide continuous coverage over large areas, making them well-suited for high-latency-tolerant services, such as global broadcasting and long-term environmental monitoring.

Nonetheless, several challenges hinder space–ground cooperation. Propagation delay and Doppler-induced waveform instability degrade transmission quality, as addressed in \cite{9852292} by a hybrid analog–digital transmitter architecture mitigating beam squint in massive MIMO LEO systems. Heterogeneous channel conditions and dynamic user demands further complicate resource allocation, motivating interference-aware opportunistic scheduling schemes \cite{10506334}. Additionally, mobility and frequent changes in topology make handover management more complex. \cite{10118845} introduced a game-based vertical handover strategy, wherein LEO satellites facilitate dynamic resource allocation, establishing a foundation for predictive mobility in SAGIN that supports ISAC.


\paragraph{Collaboration of Air and Ground} 

In the air-ground collaboration model, drones, airships, and high-altitude platforms (HAPs) serve as intermediaries between satellites and terrestrial networks, providing flexible regional coverage and rapid deployment. HAPs at 20–50 km altitude provide quasi-stationary regional coverage, while UAVs operating from 100 m to 20 km altitude offer localized and highly adaptive service. Large UAVs (around 3,000 m) are often deployed for wide-area monitoring, whereas medium-altitude UAVs (around 1,000 m) are more effective in dense urban environments. Air–ground (AG)-ISAC systems leverage aerial mobility to enhance terrestrial perception. However, they introduce synchronization, trajectory planning, and cooperation challenges. For instance, UAV trajectory optimization was shown in \cite{10049809} to reduce sensing interruptions and improve reliability, while  \cite{10614082} emphasized deep cooperation strategies for cross-layer resource coordination between aerial and terrestrial nodes.


\paragraph{Collaboration of Space and Air}

Space-air collaboration is facilitated between satellites and high-mobility aerial platforms, such as UAVs and HAPs, to extend connectivity and sensing in dynamic, dispersed environments. 
UAVs and HAPs adjust their altitudes and positions to optimize LoS with satellites, ensuring reliable, continuous data transmission. This flexibility is critical for maritime communications, remote IoT deployments, and scenarios where maintaining LoS is challenging. Moreover, this cooperation also empowers situational awareness in dynamic environments such as disasters and urban areas. 
Their high mobility and adaptability enable targeted data collection, complementing satellite and ground systems to fill coverage gaps, where conventional infrastructure is compromised or absent. The space-air collaboration layer enhances perception by enabling real-time, adaptive sensor positioning for optimal data acquisition in remote or rapidly changing areas.


\paragraph{Summary and Discussion} 

SAGIN represents a next-generation framework for extending sensing resolution, coverage, and reliability across multi-layered networks. Two main deployment paradigms have been explored. The first leverages existing terrestrial infrastructures below 300 meters altitude, offering cost efficiency but limited vertical coverage and spectrum contention between uplink, downlink, and sensing functions. The second relies on dedicated low-altitude networks above 300 m, where mmWave-based fusion architectures offer high-throughput communication and fine-grained sensing capabilities tailored to specific mission requirements. 

Despite its promise, SAGIN introduces unique technical hurdles. Spectrum sharing across segments is complicated by rapidly varying interference conditions, which necessitate collaborative spectrum sensing approaches, such as the distributed cognitive satellite–ground framework proposed in \cite{hu2020energy}. Latency and real-time data processing pose additional challenges in high-mobility aerial scenarios, where distributed edge computing and predictive beamforming techniques \cite{10745905} enable low-latency and mobility-resilient service continuity. Moreover, transmitting sensitive multi-segment data raises security and privacy concerns. Blockchain-based schemes have been proposed to guarantee tamper-proof, decentralized data integrity across space, air, and ground domains.

\begin{figure*}[htp!]
        \centering
        \includegraphics[width=0.99\linewidth]{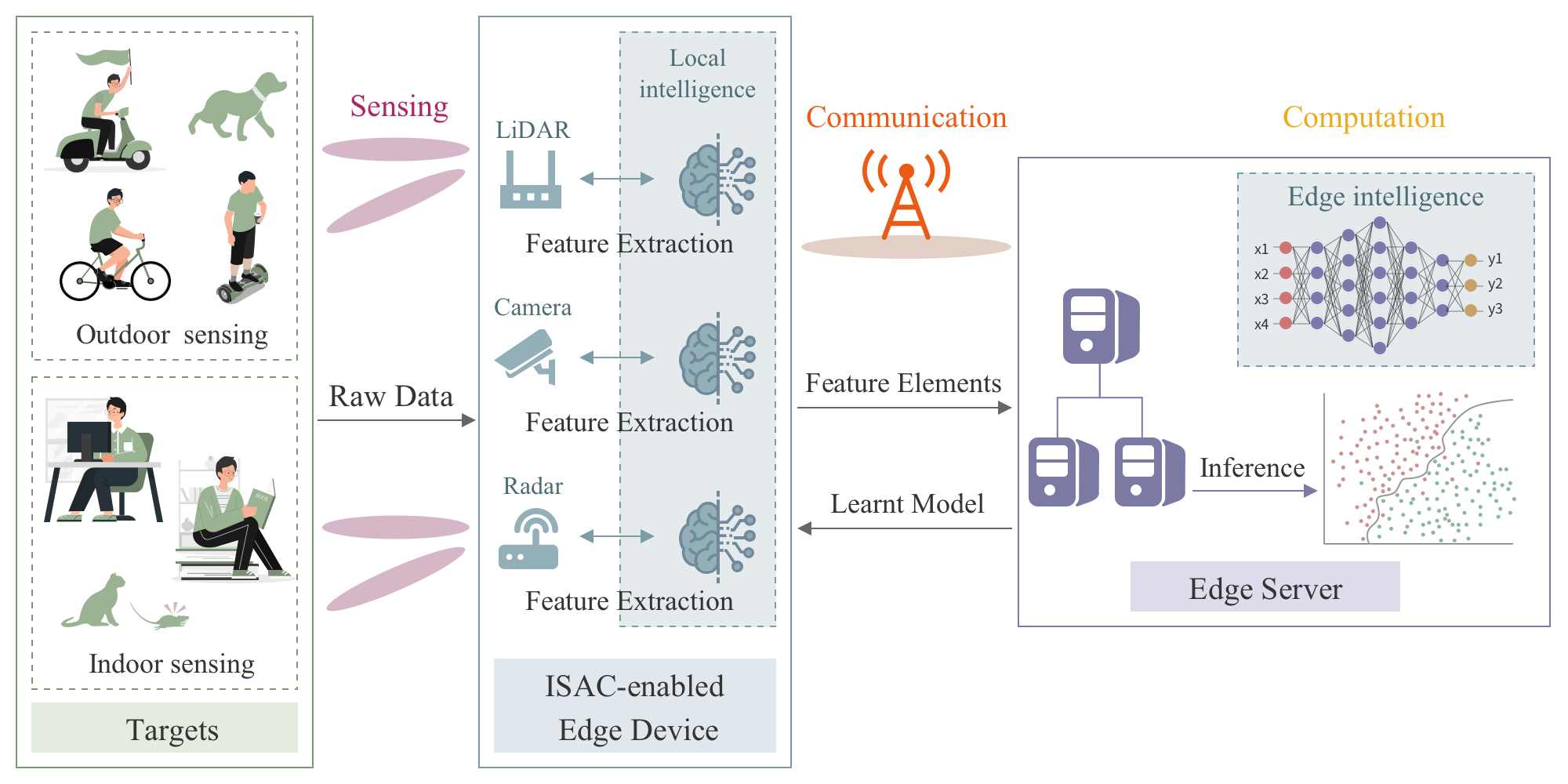}
        \caption{Illustration of the system architecture for the edge perception}
        \label{fig: Edge intelligence}
\end{figure*}

\subsection{Lessons Learned}

The transition from single-cell to multi-cell and eventually to SAGIN-based ISAC architectures illustrates a clear trajectory from localized integration toward large-scale collaborative networks. Single-cell deployments demonstrated the feasibility of waveform co-design but suffered from limited coverage, scalability, and interference. Multi-cell architectures addressed these issues by introducing BS coordination, macro–micro cooperation, and centralized C-RAN structures, which improved spatial diversity and sensing accuracy. Extending ISAC into the space–air–ground domain further enabled global applications such as navigation, disaster response, and environmental monitoring.

While these developments highlighted the potential of ISAC, they also introduced challenges in synchronization, inter-cell interference management, and coordination overhead. To address these, advanced waveform design, coordinated beamforming, and cross-layer frameworks are required to balance S\&C demands in dynamic environments.

\subsubsection{How Can Clock Asynchronization Be Effectively Mitigated in Distributed ISAC Networks}


Clock asynchronization remains a fundamental limitation in large-scale ISAC deployments. Misaligned oscillators introduce timing offsets, carrier frequency shifts, and phase noise, thereby degrading range–Doppler estimation and symbol detection. Research has investigated mitigation through CSI ratio-based models \cite{10678871}, CRB-based performance analysis \cite{9617146}, and advanced estimators such as iterative cross-correlation \cite{10273396}, Kalman filtering \cite{hamilton2008aces}, and optimization-based offset estimation \cite{10130780}. Practical implementations, such as the SPACE framework for joint synchronization and positioning and anchor-assisted uplink sensing \cite{10637442}, have integrated offset estimation and compensation. Despite these advances, scalable synchronization solutions for dynamic, multi-node ISAC networks remain limited, and trade-off analyses, such as those between synchronization accuracy and resource utilization, need further development.

\subsubsection{How Should Sensing Architectures Be Designed to Balance Latency, Fusion Accuracy, and Scalability}

Efficient sensing architecture design has been recognized as a critical enabler for the evolution of ISAC networks toward multi-cell and hierarchical deployments. At the 3GPP 6G workshop in March 2025 \cite{6gworkshop}, several candidate solutions were introduced, with a three-layer sensing framework emerging as one of the most intuitive options. This framework consists of three key components: (i) local sensing nodes, which include Remote Radio Units (RRUs), roadside units, and ISAC-enabled user equipment (UEs); (ii) intermediate aggregators, such as edge servers and distributed Baseband Units (BBUs); and (iii) the sensing function implemented within the core network. In this architecture, environmental signals are captured and pre-processed at the local nodes. The intermediate aggregators then collect and refine the measurements from multiple sources. Finally, the core network enforces system-wide requirements and carries out advanced data fusion and analytics. By distributing initial processing and partial fusion to these intermediate layers—rather than transmitting raw sensor data directly to the core—this hierarchical design reduces transmission latency, alleviates computational burdens on the sensing nodes, and mitigates the delays associated with fully centralized processing.

Looking forward, the growing density of cellular infrastructures—including terrestrial networks, vertical industries, and aerial platforms—presents a remarkable opportunity to establish ubiquitous sensing as a fundamental capability of next-generation wireless systems. By leveraging shared infrastructure, reusing spectrum, and harnessing economies of scale, ISAC can offer large-scale, cost-efficient, and resilient sensing services.

\begin{figure*}[htp!]
    \centering
    \captionsetup{justification=raggedright}  
    \includegraphics[width=0.99\linewidth]{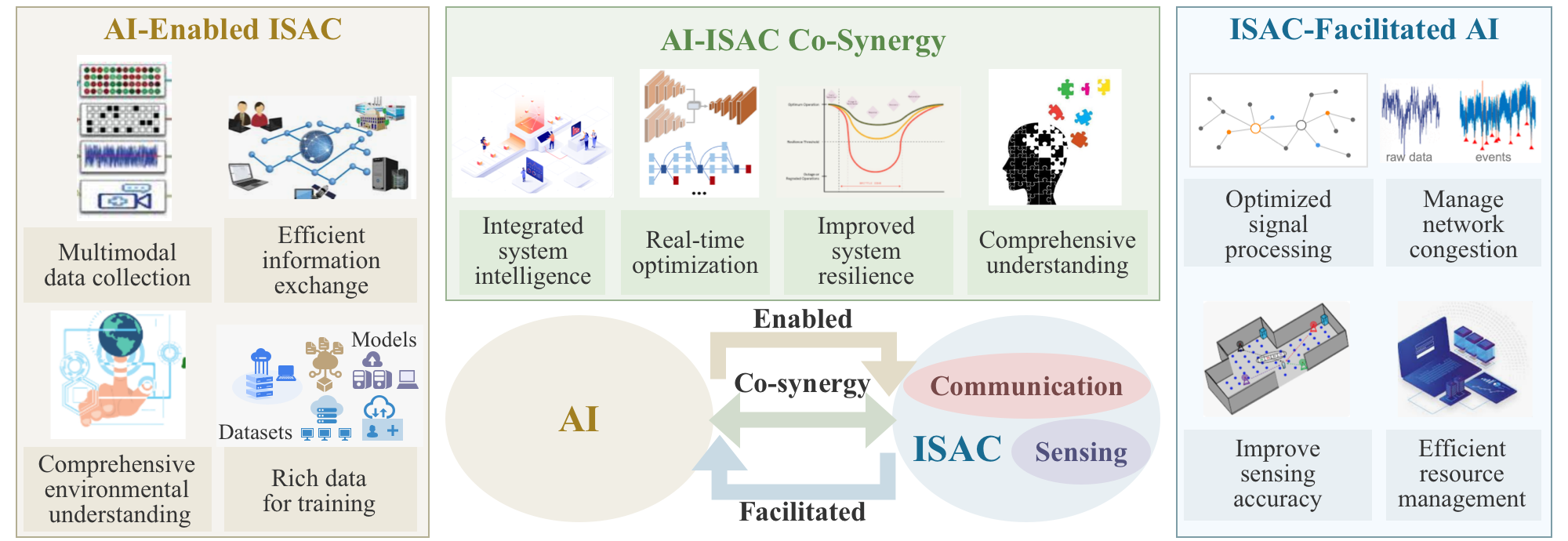}
    \caption{Interactions between AI and ISAC: AI-enabled ISAC, where AI enhances ISAC through multi-modal data fusion, efficient information exchange, and environmental understanding; ISAC-facilitated AI, in which ISAC provides precise sensing data and efficient resource management, significantly improving AI-driven signal processing and network performance; AI-ISAC Co-synergy, where AI and ISAC systems mutually reinforce each other.}
    \label{fig:Interactions between AI and ISAC}
\end{figure*}

\section{Sensing Method from Single-Modal to Multi-Modal (Edge Perception)}\label{section4}

The transition from single-modal to multi-modal approaches is pivotal in the evolution of ISAC systems, driven by the need for high-precision, real-time decision-making in applications such as autonomous driving, smart cities, and industrial automation. 
Multi-modal ISAC systems integrate heterogeneous data sources to enhance both environmental perception and communication performance. Edge intelligence accelerates this shift by enabling decentralized, real-time data processing at the network edge, reducing communication overhead, and addressing privacy concerns.

This section reviews the broadening of ISAC technologies from single-modal architectures to multi-modal integration, with particular emphasis on the interplay between ISAC and AI. By leveraging AI-driven techniques, both S\&C functionalities are significantly enhanced to meet the requirements of next-generation intelligent applications.


\subsection{From Edge Intelligence to Edge Perception}


Modern smart environments are supported by billions of edge devices (e.g., smartphones, wearable sensors, and connected vehicles) that generate massive volumes of heterogeneous data. Traditional cloud-centric approaches, which offload computation to centralized servers, often incur high latency, excessive communication overhead, and elevated privacy risks \cite{9815195}. Edge intelligence addresses these limitations by distributing computation across edge nodes, enabling localized data analysis and decision-making\cite{LiTWC2020,Gartner}. Federated learning (FL), in particular, offers a privacy-preserving solution by training models locally while aggregating global insights at edge servers\cite{Cao-ComM, 10478029}.

Despite its advantages, edge intelligence faces persistent challenges. The transmission of large models and gradients can overload wireless networks, and limited channel quality (e.g., fading) impairs global model convergence. Device heterogeneity further complicates federated training, while adversarial participants may exploit vulnerabilities to manipulate sensor inputs.

\subsubsection{Emergence of Edge Perception}

The integration of sensing into edge intelligence gives rise to edge perception, where data are processed closer to end devices for responsive and privacy-preserving analytics \cite{10908560}. As shown in Fig. \ref{fig: Edge intelligence}, an edge perception setup typically includes three co-located modules: (i) a sensing module capturing environmental data, (ii) a computing module extracting salient features, and (iii) a communication module transmitting features to remote servers. This localized decision-making reduces latency and alleviates the burden on backhaul networks.

\subsubsection{Edge Perception for Sensing Enhancement}

Edge perception plays a central role in enhancing sensing capabilities through multi-modal data fusion at the network edge. By employing advanced AI algorithms, subtle patterns that elude conventional methods can be captured, thereby improving detection accuracy and situational awareness. For instance, in smart homes, sensor data are monitored to assess occupant behavior and environmental factors, which in turn facilitate the automated optimization of temperature, humidity, and lighting to enhance comfort. In industrial IoT, real-time factory monitoring is utilized to support predictive maintenance and production scheduling, while wearable devices track patient vitals to enable prompt medical interventions. Beyond indoor and industrial settings, edge perception also plays a critical role in highly dynamic outdoor environments. In dense low-altitude operations, graph neural network–based methods enable UAVs to model local state information, extract multi-modal sensing features for conflict-risk awareness, and improve the effectiveness of autonomous avoidance decisions in large-scale swarms~\cite{11159205,11146533}.

External sensing is conducted by edge perception systems that are equipped with on-device processing capabilities. Once sensor data are generated at distributed sources, they are gathered and analyzed locally using intelligent algorithms, which enables swift decision-making while conserving network bandwidth. 
Internal sensing, on the other hand, involves the continuous monitoring of network conditions (such as congestion and bandwidth constraints) to support responsive scheduling and resource allocation. For example, in intelligent transportation systems, traffic signals are dynamically reconfigured based on real-time congestion indicators, and video surveillance frameworks adjust streaming quality to maintain the integrity of critical feeds under limited bandwidth conditions.

\begin{figure*}[!htp]
    \centering
    \includegraphics[width=0.99\linewidth]{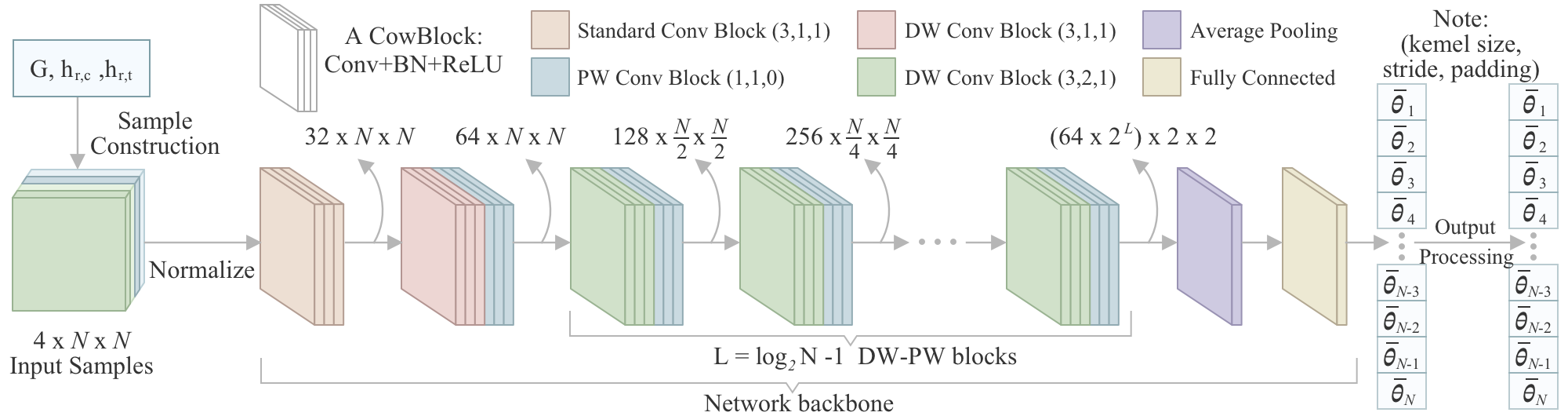}
    \caption{Illustrate the structure of the ISAC beamforming neural network in paper\cite{10533223}}
    \label{fig:ibfnet}
\end{figure*}

\subsubsection{Edge Perception for Communication Enhancement}

In addition to sensing, edge perception strengthens communication performance by fusing diverse sensing modalities into the transmission process. Localized processing reduces end-to-end latency and alleviates computational pressure on centralized infrastructure. Multi-modal fusion improves communication reliability and resource management. In vehicular networks, radar, camera, and LiDAR inputs can be combined with RF signals to enhance interference mitigation, beamforming precision, and spectrum allocation. This synergy increases throughput and reduces latency. Beyond resource optimization, reinforcement learning and spiking neural networks dynamically allocate spectrum resources, further improving adaptability and robustness in highly dynamic environments \cite{shang2025energyefficientintelligentisacv2x}.

\subsubsection{Interplay between ISAC and AI}

The integration of ISAC and AI is inherently bidirectional, with each domain augmenting the other. As shown in Fig. \ref{fig:Interactions between AI and ISAC}, this interplay can be understood through three key perspectives.

\paragraph{Edge AI-Enabled ISAC}


Edge computing provides rapid model adaptation and low-latency inference for ISAC, particularly on resource-constrained devices. Unlike federated learning, which emphasizes distributed training, practical deployments prioritize lightweight edge-side inference and fine-tuning of pretrained models at the edge side.

Techniques such as neural network compression, shallow architectures, and meta-learning have been adopted to support low-overhead inference. For instance, Ahmed et al. \cite{9467318} proposed a shallow neural network for hybrid beam selection, achieving real-time operation with a 40\% reduction in computational cost while maintaining near-optimal performance in mmWave MIMO systems. Similarly, Yuan et al. \cite{9257198} introduced a meta-learned beamforming strategy capable of rapid adaptation to new propagation scenarios with limited training samples, greatly accelerating the convergence process. Transfer learning has further enabled robust generalization, where pretrained policies in \cite{9367008} achieved performance comparable to iterative baselines while substantially reducing the computational complexity required for reconfiguration.

These findings highlight the potential of pretrained AI models, once optimized for edge constraints, to be efficiently fine-tuned or directly deployed in ISAC systems requiring responsiveness, adaptability, and low computational complexity.



\paragraph{ISAC-Facilitated Edge AI}


The integration of S\&C within ISAC frameworks provides a synchronized, multimodal data source that significantly enhances edge AI inference. Unlike traditional architectures relying on disjointed sensor modalities, ISAC systems offer a unified and temporally aligned representation of environmental dynamics by fusing radar, LiDAR, vision, and RF features.

This synergy has been validated across multiple application domains. In vehicular networks,  a transformer-based predictive beamforming scheme that leveraged radar echoes to bypass explicit CSI acquisition improved beam selection accuracy and link throughput, achieving substantial gains even under high-mobility conditions \cite{zhang2024transformer}. In indoor sensing, vision transformer–based models effectively processed Wi-Fi CSI data for human activity recognition (HAR). In particular, the class-attention in image transformers architecture achieved state-of-the-art classification accuracy on the University of Texas Human Activity Recognition (UT-HAR) dataset under constrained edge inference conditions \cite{10477406}. Similarly, the ImgFi framework transforms CSI into image representations to leverage 2D CNN-based architectures, achieving up to 99.5\% recognition accuracy on benchmark datasets, with a lightweight classification model requiring approximately 0.1~GFLOPS of computation and 6.5~MB of parameters~\cite{10190332}.

\begin{table*}[htp!]
\centering
\caption{Summary of Deep Learning Methods of ISAC systems}
\label{tab:Deep Learning Methods}
\begin{tblr}{
  width = \linewidth,
  colspec = {Q[60]Q[190]Q[130]Q[255]Q[270]},
  row{1} = {c},
  cell{2}{1} = {c},
  cell{2}{3} = {r=2}{c},
  cell{3}{1} = {c},
  cell{4}{1} = {c},
  cell{4}{3} = {c},
  cell{5}{1} = {c},
  cell{5}{3} = {c},
  cell{6}{1} = {c},
  cell{6}{3} = {r=2}{c},
  cell{7}{1} = {c},
  cell{8}{1} = {c},
  cell{8}{3} = {r=3}{c},
  cell{9}{1} = {c},
  cell{10}{1} = {c},
  cell{11}{1} = {c},
  cell{11}{3} = {r=3}{c},
  cell{12}{1} = {c},
  cell{13}{1} = {c},
  cell{14}{1} = {c},
  cell{14}{3} = {c},
  cell{15}{1} = {c},
  cell{15}{3} = {r=2}{c},
  cell{16}{1} = {c},
  cell{17}{1} = {c},
  cell{17}{3} = {r=2}{c},
  cell{18}{1} = {c},
  cell{19}{1} = {c},
  cell{19}{3} = {r=2}{c},
  cell{20}{1} = {c},
  vlines,
  hline{1-2,4-6,8,11,14-15,17,19,21} = {-}{},
  hline{3,7,9-10,12-13,16,18,20} = {1-2,4-5}{},
}
\textbf{Reference}   & \textbf{Signal Processing}                                                  & \textbf{Method}              & \textbf{Application}                                                            & \textbf{Performance}                                                                                   \\
\cite{9937293}              & Velocity-time map extraction via range FFT and Doppler FFT, noise filtering & CNN                          & On-edge human activity classification (mmWave radar sensing)                    & Subject-independent accuracy 96.43\%, inference latency 120 ms, model size 1.44 KB                     \\
\cite{9931675 }             & Low-pass filtering with Gaussian smoothing                                  &                              & Hand Gesture Recognition                                                        & Average accuracy 96.31\%, F1-score 0.97                                                                \\
\cite{9720163 }             & Sparse point cloud processing                                               & Message Passing Neural Network/ Graph Convolution          & Gesture recognition                                                             & Real-time accuracy 90.53\%, up to 98.7\% on standard datasets                                                                                          \\
\cite{10606321 }            & Matched filtering on radar echoes                                           & Bidirectional LSTM-based RNN & Joint vehicle association and predictive beamforming in vehicular ISAC networks & Throughput under arrival rate 12.5 vehicles/s                                                   \\
\cite{grobelny2022mm}       & Fast fourier transform (FFT) for radar signal processing, noise filtering   & LSTM                         & Hand gesture recognition for contactless human-machine interfaces               & Accuracy 94.4\%, accuracy drop $\le$2.24\% for unseen subjects        \\
\cite{10571266 }            & Pre-processing with normalization and filtering                             &                              & Autonomous driving, smart cities                                                & Detection probability 80\%, RMSE reduction 23.5\%                                                      \\
\cite{9906941    }          & Processing of angle data with noise filtering                               & CNN+LSTM                     & Predictive beamforming in vehicular networks                                    & \textasciitilde{}5 dB sum-rate gain over model-based methods, NMSE = 0.7                               \\
\cite{huang2022activity}    & Cell average constant false-alarm rate algorithm, data filtering            &                              & Human activity recognition (boxing, jumping, squatting, etc.)                   & Recognition accuracy 97.26\%, 1\% higher than using single data types                                     \\
\cite{hayashi2021radarnet}  & Complex range--Doppler map extraction via FFT-based radar signal processing &                              & Gesture recognition                                                             & Accuracy > 99\% in segmented classification, inference time 0.147\textasciitilde{}ms \\
\cite{zhang2024transformer} & Echo signal processing and angle feature sequence construction              & Transformer                  & Predictive beamforming for vehicular ISAC systems                               & Sum-rate gain 3.7 dB over baseline under high mobility                                                 \\
\cite{10477406  }           & CSI denoising, PCA, STFT-based spectrogram generation                       &                              & Human activity recognition (Wi-Fi sensing)                                      & Accuracy 98\%~                                                                                         \\
\cite{10577431 }            & Position-based data normalization, LiDAR downsampling                       &                              & Beam prediction in mmWave V2I ISAC systems                                      & Distance-based accuracy 91\%, improved zero-shot robustness                                            \\
\cite{9848800}              & Heartbeat micro-Doppler extraction, band-pass filtering, phase unwrapping   & Supervised Learning          & Authentication (Heart rate data)                                                & Average accuracy >95\% \\
\cite{10533223  }           & Channel correlation and sensing gain maximization                           & Unsupervised Learning        & Beamforming design in RIS-aided ISAC                                            & Sensing SNR $\approx$ 11--13 dB (near-optimal)
                                            \\
\cite{10113889 }            & Local CSI exchange, interference power measurement, normalization           &                              & Interference management in multi-cell ISAC systems                              & Interference mitigation 30\%, reduced overhead by 20\%~                                                \\
\cite{10621049  }           & Over-the-air FL  aggregation, beamforming                                    & Federated Learning           & ISAC  networks integrating over-the-air FL for model training                   & Aggregation latency independent of device number, radar SNR $\geq 15$ dB
                                     \\
\cite{10669033    }         & GAN-based data  augmentation, deep Q-Network Optimization                   &                              & Large-Scale Multi-agent ISAC                                                    & Reduces  communication cost by 35\%, improves FL convergence by 1.6x                                   \\
\cite{10620899  }           & UAV trajectory discretization, RIS phase shift control                      & Multi-agent Learning         & UAV  RIS-assisted ISAC optimization                                             & Achieves higher sensing rate and communication rate than non-learning baselines                        \\
\cite{10571018  }           & Matched filtering, CRLB-based sensing modeling                              &                              & Joint beamforming and power allocation in ISAC-assisted V2I networks            & Sum-rate $\approx$7.5 bit/s/Hz, angle CRLB $\approx10^{-4}$ rad, distance CRLB $\approx10^{-7}$ m
   
\end{tblr}
\end{table*}

Beyond these applications, ISAC-enabled multimodal observation also strengthens multi-agent decision-making in highly dynamic environments. In UAV swarms, multi-agent deep reinforcement learning methods based on spatiotemporal graph reasoning and transformer-based sequence modeling enable efficient edge-node communication and coordination, significantly improving cooperative pursuit success probabilities against highly maneuverable targets under uncertain environments~\cite{11159205,11146533,10659125}.


As pointed out, ISAC not only enhances the quality and diversity of sensing data but also enables efficient and low-latency AI inference at the edge, thereby supporting scalable deployment in latency-sensitive applications.


\paragraph{AI–ISAC Co-Synergy}

The interaction between AI and ISAC is inherently bidirectional, with each component reinforcing the capabilities of the other. On one hand, AI algorithms can be employed to optimize waveform selection, beamforming, and sensing strategies in real time. On the other hand, ISAC systems provide rich, temporally aligned environmental feedback, enabling AI to refine inference and decision-making processes with improved accuracy and robustness.

This co-synergistic paradigm has been exemplified in \cite{10621049}, where an ISAC-guided over-the-air federated learning framework was introduced. By leveraging sensing data quality for dynamic client scheduling and power control, the framework improved convergence robustness and reduced training latency in bandwidth-limited environments\cite{11143883}. To further enhance adaptability, \cite{10608156} proposed a transfer learning–based edge model adaptation framework, which enabled fine-tuning of pretrained models using few-shot local samples. This design preserved model accuracy and privacy while significantly lowering communication overhead, demonstrating the potential of ISAC-assisted adaptive learning in dynamic wireless contexts.

Looking ahead, co-optimized ISAC–AI architectures are expected to incorporate continual learning, meta-learning, and online adaptation mechanisms that dynamically update AI models in response to real-time feedback. Such integration advances the resilience and efficiency of edge intelligence and further establishes ISAC as a fundamental enabler of self-evolving, autonomous wireless systems.




\paragraph{Challenges and Limitations}


Despite the promising synergy, the deployment of AI-driven ISAC systems remains constrained by multiple technical challenges.


A primary concern lies in the requirement for precise time and frequency synchronization across distributed S\&C nodes, particularly under dynamic and mobile environments. Clock and carrier frequency offsets have been shown to significantly degrade model aggregation in federated learning and reduce the accuracy of over-the-air computation. To address this issue, distributed MIMO sensing under timing and frequency asynchronization was investigated in \cite{10130780}, where a robust localization algorithm was proposed to mitigate synchronization-induced degradation.


Another critical challenge arises from the instability of AI algorithms in non-stationary wireless environments. RL methods commonly employed for optimizing beamforming and waveform design often rely on stationary assumptions that fail to generalize under rapid channel variations and user mobility. To improve robustness, a DRL-based beamforming strategy was developed in \cite{10571018}, incorporating mobility and environmental dynamics to enhance performance in vehicular scenarios.


Energy efficiency and sensing cost further represent significant barriers in resource-constrained deployments. To alleviate these constraints, a closed-loop AI-in-the-loop framework was introduced in \cite{cai2025aiintheloopsensingcommunicationjoint}, where gradient importance sampling and stochastic gradient Langevin dynamics (SGLD)-based noise modeling were applied, achieving a 52\% reduction in sensing cost and a 77\% saving in communication energy. In passive sensing scenarios, a deep unfolding architecture named ISAC-NET was proposed in \cite{10474422}, which alternates between channel reconstruction and target detection, leading to consistently improved sensing accuracy and channel estimation performance compared with conventional methods.


\subsection{The Goal of the Next Phase}

The development of next-generation networks aims to integrate communication, sensing, and computing into unified cognitive infrastructures that augment human intelligence. In current distributed AI frameworks, edge nodes already provide essential capabilities such as sensing, autonomy, real-time responsiveness, and adaptive learning. The long-term objective is to move beyond reactive networks—primarily focused on perception and environmental understanding—toward fully autonomous cognitive systems that support decision-making, actuation, control, and continual self-optimization. 

\subsubsection{The Shift towards Task-Oriented ISAC}

 \begin{figure}[htp!]
    \centering
    \includegraphics[width=0.99\linewidth]{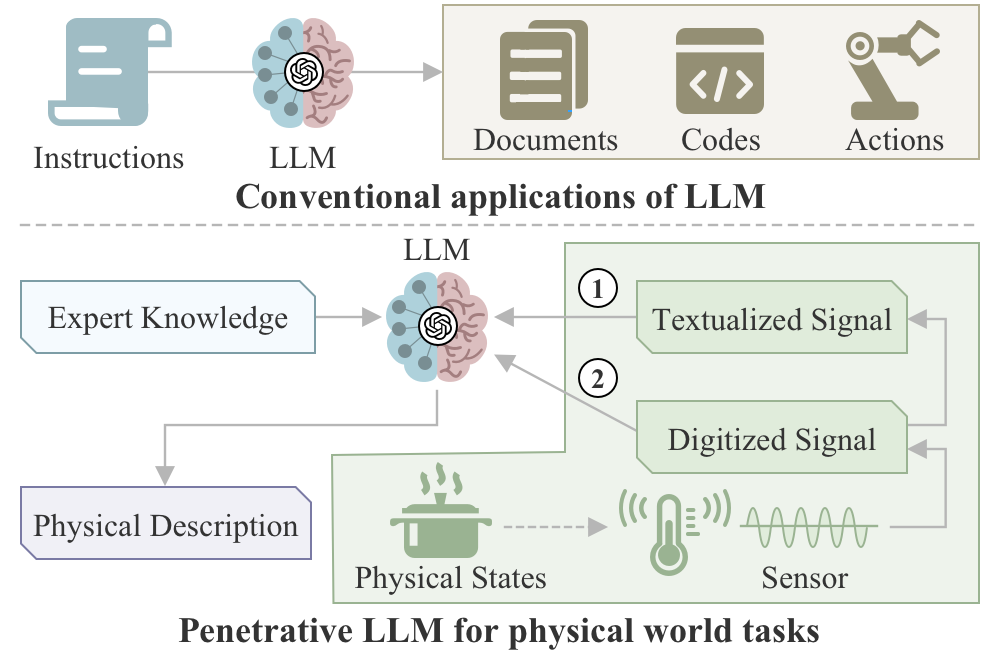}
    \caption{The structure in paper\cite{xu2024penetrative}}
    \label{fig:ecg}
\end{figure}

The integration of sensing, communication, and edge intelligence represents a fundamental paradigm shift, where the emphasis shifts from maximizing individual performance to accomplishing tasks efficiently and reliably. In future networks, vast amounts of multi-sensory data with diverse spatial, temporal, and stochastic characteristics will be generated at the edge. The processing of this data will directly affect edge intelligence and overall task execution. Moreover, the relative priority of S\&C services will vary depending on task requirements. Consequently, data collection and transmission need to be tightly coupled with edge computing, ensuring that QoS definitions for ISAC are redefined from a task-oriented perspective. Within this framework, edge devices are expected to deliver optimized sensing performance while operating under constrained communication and computing resources.



\subsubsection{Enabling AI Technologies} 


By leveraging advanced algorithms to process and interpret high-dimensional sensory data, AI transforms conventional networks into adaptive, context-aware infrastructures. For instance, a lightweight unsupervised neural network was proposed in \cite{10533223} to optimize beamforming in RIS-aided ISAC systems, achieving reduced computational complexity and enhanced spectral efficiency, as illustrated in Fig. \ref{fig:ibfnet}. Representative AI-based methods for ISAC are summarized in Table \ref{tab:Deep Learning Methods}.


Among recent breakthroughs, large language models (LLMs) have demonstrated strong potential in extracting knowledge from sensory data with minimal task-specific supervision. Typically structured into three modules, including perception, grounding, and alignment, LLMs enable multimodal reasoning and semantic abstraction\cite{10638143}. For example, \cite{xu2024penetrative} introduced a framework where smartphone accelerometer signals were transformed into textual sequences for motion and environment inference, while electrocardiogram (ECG) signals were converted into text-based indicators such as heart rate, as shown in Fig. \ref{fig:ecg}.


However, deploying LLMs within ISAC remains a challenging task due to the latency requirements, model complexity, and hardware limitations. As noted in \cite{11026879}, cloud-based deployment supports high-level semantic reasoning but suffers from excessive transmission latency, rendering it unsuitable for real-time tasks. Conversely, edge-based deployment enhances privacy and eliminates transmission overhead, but faces computational bottlenecks where processing delays for complex LLMs can exceed 50 ms. Hybrid edge–cloud strategies provide a compromise, performing prompt-based summarization at the edge while leveraging cloud resources for advanced reasoning. Mapping deployment strategies to ISAC use cases reveals that real-time tasks such as beam prediction and mobility management benefit from edge or hybrid solutions, while latency-tolerant tasks like semantic reasoning or long-term optimization prefer cloud-based solutions. KPIs include inference delay, power consumption, model size, and Top-$\kappa$ accuracy in sensing-driven tasks. For instance, \cite{10829757} demonstrated that ChatGPT-4, when guided by appropriate prompts, could effectively analyze multimodal data for beam prediction, consistently outperforming baseline algorithms in prediction accuracy across different scenarios.

Although challenges regarding interpretability, energy efficiency, and real-time constraints remain, LLM-enabled ISAC frameworks offer a promising pathway toward scalable and intelligent S\&C infrastructures in next-generation wireless networks.



\subsection{ISAC Datasets}


The design of communication modules is inherently coupled with wireless channel characteristics, which are typically analyzed using large-scale channel measurements. Similarly, the design of sensing systems relies on multimodal data collected from diverse sensors. Therefore, constructing comprehensive sensing datasets is of critical importance for advancing edge intelligence research and enabling the effective application of AI in ISAC.


Inspired by ImageNet \cite{5206848}, recent efforts have proposed systematic methodologies for building large-scale sensing datasets. The process generally involves four key steps: 1) Defining dataset objectives to guide categorization and ensure diversity; 2) Collecting massive multimodal data from heterogeneous sources such as LiDAR, cameras, mobile devices, and wireless sensors; 3) Cleaning and labeling data through automated or crowdsourcing approaches with confidence thresholds and duplicate removal; 4) Making datasets publicly available via cloud platforms to foster community contributions and benchmarking.



\subsubsection{High-Quality Datasets}
Several representative datasets have been developed to support ISAC and edge intelligence research:

\begin{itemize}

\item SDP\cite{sdp8}: Launched in February 2024 by IEEE ISAC-ETI, the SDP platform is the industry's first large-scale field-measured sensing dataset. This platform addresses major wireless sensing challenges by publicly releasing RF datasets and promoting high-performance and reliable wireless sensing solutions. The platform has released a large-scale field-measured dataset that covers 7 typical sensing application scenarios. It spans 400 hours and can be used for algorithm research and model training for various sensing tasks. These tasks include detection (presence detection of targets like people and pets), positioning \& tracking (tracking of people, vehicles, and drones), recognition (gesture and action recognition), vital signs (breathing, heartbeat, sleep), and imaging (human skeleton and environmental reconstruction).

\item WiMANS\cite{huang2025wimans}: A Wi-Fi-based multi-user activity sensing benchmark, WiMANS provides 9.4 hours of dual-band CSI synchronized with RGB videos of up to five users performing nine daily activities. It targets multi-user activity recognition, tracking, and pose estimation, addressing the limitations of single-user datasets.

\item Radar signatures of human activities\cite{fioranelli2019radar}: This dataset, provided by the University of Glasgow, focuses on radar-based human activity recognition. It includes a range of human activity radar signals collected using FMCW radar, supporting the development and evaluation of radar signal processing and machine learning algorithms for human activity classification.

\item EyeFi\cite{9183685}: EyeFi fuses CSI with camera data for human sensing, motion tracking, and identification tasks. It uses a three-antenna Wi-Fi chipset to estimate the Angle of Arrival (AoA) of Wi-Fi signals combined with overhead camera trajectories. Synchronization aligns CSI-based and vision-based trajectories. EyeFi performs trajectory matching to identify individuals by linking their Wi-Fi MAC addresses and physical movements.

\item OPERAnet\cite{bocus2022operanet}: OPERAnet is a multi-modal dataset designed for indoor Human Activity Recognition (HAR) and localization. The dataset contains approximately 8 hours of data collected across two indoor environments with six participants performing six daily activities. It is designed to facilitate multi-modal research in smart homes, elderly care, and surveillance, with high-resolution activity and location labels.

\item MmWave Gesture Dataset\cite{liu2021m}: MmWave-gesture-dataset features over 54,620 gesture instances and 1,357 minutes of gesture data, collected from 15 scenarios with varying distance ranges. It includes both preset and unexpected gestures, offering multi-level mmWave data ranging from raw signals and Range-Doppler Images (RDIs) to radar-derived surface energy point (SEP) cloud representations, which aids in gesture recognition technology development.

\begin{table*}
\centering
\caption{Summary of High-Quality Datasets in ISAC}
\label{tab:Dataset}
\begin{tblr}{
  width = \linewidth,
  colspec = {Q[108]Q[155]Q[80]Q[80]Q[80]Q[90]Q[80]Q[105]Q[135]},
  row{1} = {c},
  row{2} = {c},
  cell{1}{1} = {r=2}{},
  cell{1}{2} = {r=2}{},
  cell{1}{3} = {c=5}{0.48\linewidth},
  cell{1}{8} = {r=2}{},
  cell{1}{9} = {r=2}{},
  cell{3}{1} = {c},
  cell{3}{3} = {c},
  cell{3}{4} = {c},
  cell{3}{5} = {c},
  cell{3}{6} = {c},
  cell{3}{7} = {c},
  cell{4}{1} = {c},
  cell{4}{3} = {c},
  cell{4}{4} = {c},
  cell{4}{5} = {c},
  cell{4}{6} = {c},
  cell{4}{7} = {c},
  cell{5}{1} = {c},
  cell{5}{3} = {c},
  cell{5}{4} = {c},
  cell{5}{5} = {c},
  cell{5}{6} = {c},
  cell{5}{7} = {c},
  cell{5}{8} = {c},
  cell{6}{1} = {c},
  cell{6}{3} = {c},
  cell{6}{4} = {c},
  cell{6}{5} = {c},
  cell{6}{6} = {c},
  cell{6}{7} = {c},
  cell{6}{8} = {c},
  cell{7}{1} = {c},
  cell{7}{3} = {c},
  cell{7}{4} = {c},
  cell{7}{5} = {c},
  cell{7}{6} = {c},
  cell{7}{7} = {c},
  cell{8}{1} = {c},
  cell{8}{3} = {c},
  cell{8}{4} = {c},
  cell{8}{5} = {c},
  cell{8}{6} = {c},
  cell{8}{7} = {c},
  cell{9}{1} = {c},
  cell{9}{3} = {c},
  cell{9}{4} = {c},
  cell{9}{5} = {c},
  cell{9}{6} = {c},
  cell{9}{7} = {c},
  cell{10}{1} = {c},
  cell{10}{3} = {c},
  cell{10}{4} = {c},
  cell{10}{5} = {c},
  cell{10}{6} = {c},
  cell{10}{7} = {c},
  cell{10}{8} = {c},
  cell{11}{1} = {c},
  cell{11}{3} = {c},
  cell{11}{4} = {c},
  cell{11}{5} = {c},
  cell{11}{6} = {c},
  cell{11}{7} = {c},
  vlines,
  hline{1,3-12} = {-}{},
  hline{2} = {3-7}{},
}
\textbf{Dataset Name}                & \textbf{Data Types} & \textbf{Application Scenarios} &                      &                   &                      &                  & \textbf{Data Volume}                            & \textbf{Provider/Founder}                         \\
& & \textbf{Detection}             & \textbf{Positioning} & \textbf{Tracking} & \textbf{Recognition} & \textbf{Imaging} &                                                 &                                                   \\
SDP\cite{sdp8}                                  & {\labelitemi\hspace{\dimexpr\labelsep+0.5\tabcolsep}CSI 
\\\labelitemi\hspace{\dimexpr\labelsep+0.5\tabcolsep}Kinect data} & \checkmark  &  \checkmark  &  \checkmark   &   \checkmark &  \checkmark & {Time >400 hours}               & IEEE ISAC-ETI                                     \\
WiMANS\cite{huang2025wimans}                               & {\labelitemi\hspace{\dimexpr\labelsep+0.5\tabcolsep}CSI
\\\labelitemi\hspace{\dimexpr\labelsep+0.5\tabcolsep}RGB video} & $\times$  & $\times$  &  \checkmark  & \checkmark  &  $\times$ & Time >9.4 hours                                 & Imperial College London                           \\
Radar Signatures of Human Activities\cite{fioranelli2019radar} & \labelitemi\hspace{\dimexpr\labelsep+0.5\tabcolsep}Radar& $\times$  & $\times$  & $\times$  & \checkmark  &  $\times$  &  - & University of Glasgow                             \\
EyeFi\cite{9183685}                                & {\labelitemi\hspace{\dimexpr\labelsep+0.5\tabcolsep}CSI
\\\labelitemi\hspace{\dimexpr\labelsep+0.5\tabcolsep}RGB video} & $\times$  & \checkmark  & \checkmark  & \checkmark  &  $\times$ & - & University of North Carolina                      \\
OPERAnet\cite{bocus2022operanet}                             & {\labelitemi\hspace{\dimexpr\labelsep+0.5\tabcolsep}CSI
\\\labelitemi\hspace{\dimexpr\labelsep+0.5\tabcolsep}Kinect data} & $\times$  & \checkmark  & $\times$  & \checkmark  &  $\times$ &  Time >8 hours                                   & Bocus et al.                                      \\
MmWave Gesture Dataset\cite{liu2021m}               & {\labelitemi\hspace{\dimexpr\labelsep+0.5\tabcolsep}mmWave
\\\labelitemi\hspace{\dimexpr\labelsep+0.5\tabcolsep}Raw RDIs
\\\labelitemi\hspace{\dimexpr\labelsep+0.5\tabcolsep}SEP's cloud points} & $\times$  & $\times$  & $\times$  & \checkmark  &  $\times$ &  {Size >54,620
\\Time >1357 minutes} & Beijing University of Posts and Telecommunications \\
DeepSense 6G \cite{10144504}  & {\labelitemi\hspace{\dimexpr\labelsep+0.5\tabcolsep}RGB
\\\labelitemi\hspace{\dimexpr\labelsep+0.5\tabcolsep}mmWave
\\\labelitemi\hspace{\dimexpr\labelsep+0.5\tabcolsep}LiDAR
\\\labelitemi\hspace{\dimexpr\labelsep+0.5\tabcolsep}GPS
\\\labelitemi\hspace{\dimexpr\labelsep+0.5\tabcolsep}Radar}          & \checkmark     & \checkmark     &\checkmark           & $\times$           & $\times$      & Size >1 million data points                      & Arizona State University                          \\
WALDO\cite{nist-waldo}  & {\labelitemi\hspace{\dimexpr\labelsep+0.5\tabcolsep}mmWave received signals
\\\labelitemi\hspace{\dimexpr\labelsep+0.5\tabcolsep}(optional) Channel data (for signal generation)  }                                                                           & $\times$   & \checkmark   &$\times$ &$\times$     &$\times$    & Size >850 GB (rx signals) / >395 GB (channels)
& National Institute of Standards and Technology  \\
M\textsuperscript{3}SC datasets \cite{10330577}     & {\labelitemi\hspace{\dimexpr\labelsep+0.5\tabcolsep}LiDAR
\\\labelitemi\hspace{\dimexpr\labelsep+0.5\tabcolsep}RGB
\\\labelitemi\hspace{\dimexpr\labelsep+0.5\tabcolsep}Depth maps
\\\labelitemi\hspace{\dimexpr\labelsep+0.5\tabcolsep}mmWave radar
\\\labelitemi\hspace{\dimexpr\labelsep+0.5\tabcolsep}IMU}

& $\times$& \checkmark  & \checkmark  &$\times$  &$\times$ & Size >1500 snapshots                            & School of Electronics, Peking University          
\end{tblr}
\end{table*}

\item DeepSense 6G\cite{10144504}: Designed to support research in telecommunications and sensory technologies for 6G networks, DeepSense 6G dataset provides a rich collection of real-world, multi-modal data. It captures communication and sensory inputs across diverse scenarios and times of day, facilitating deep insights into 6G technology performance and adaptability.

\item WALDO\cite{nist-waldo}: The WALDO dataset was developed for the ITU AI/ML 5G Challenge. It focuses on sensing using mmWave communications and machine learning. The dataset provides a massive collection of received signal and channel data files (available in MATLAB format) to support the development of advanced positioning algorithms. It offers two main configurations: a received signals dataset (>850 GB) and a channel data dataset (>395 GB), enabling research into accurate user localization using 5G mmWave signals.

\item M$^{3}$SC dataset\cite{10330577}: The M$^{3}$SC dataset is a simulation-based dataset constructed in a dynamic vehicular V2X crossroad scenario. It integrates aligned multi-modal data from RGB images, depth maps, LiDAR, mmWave radar, and wireless communication channels. The dataset contains 1,500 snapshots collected under varying weather conditions (sunny, rainy, and snowy) and lighting settings (day and night). It facilitates research on intelligent multi-modal sensing--communication integration, including environment-aware wireless modeling and learning-based joint sensing and communication design.

\end{itemize}

\subsubsection{Summary and Discussion} Simulation datasets provide controllable environments for analyzing system performance and optimizing algorithm design. However, they often fail to capture the complexity and dynamics of real-world environments. By contrast, field-measured datasets such as SDP and DeepSense 6G, though resource-intensive to construct, provide higher fidelity, diversity, and representativeness. Such datasets are indispensable for training AI models that generalize well to practical ISAC scenarios, thereby accelerating progress in ISAC and edge intelligence technologies\cite{wang2025surveywifisensinggeneralizability}.

\subsection{Lessons Learned}


Despite notable application-driven advances, substantial theoretical and practical challenges persist in unifying edge intelligence with ISAC. This raises several fundamental questions:

\subsubsection{How to Build Hardware That Unifies Sensing, Communication, and Computing}


Current BS architectures are inherently complex, and the direct integration of sensing functionalities is neither cost-efficient nor scalable. Sensing modules often demand high-resolution and high-precision devices, which substantially increase implementation cost and design complexity. In parallel, AI algorithms typically rely on specialized accelerators such as GPUs and NPUs, further complicating their integration into low-power, resource-constrained communication hardware. As a result, achieving hardware platforms that seamlessly harmonize communication, sensing, and computing functionalities remains a critical frontier for future research.

\subsubsection{How Can Massive, Heterogeneous Sensing Data Be Integrated at Scale}
    

The fusion and processing of heterogeneous, multi-dimensional sensing data demand new algorithmic frameworks that can adapt to device diversity and coupled application requirements. Within a single coverage area, sensing data are inherently interdependent, necessitating efficient mechanisms for distributed model parameter generation. However, current distributed learning approaches often overlook device heterogeneity, resulting in inefficiencies and degraded performance. In massive machine-type communication (mMTC) scenarios, resource-constrained sensors may lack the capacity to perform local training, which forces trade-offs between S\&C under stringent resource budgets \cite{ullah2018information}. Developing scalable, adaptive, and resource-aware data fusion strategies remains an open and critical challenge.

    

 \begin{figure*}[htp!]
        \centering
        \includegraphics[width=0.95\linewidth]{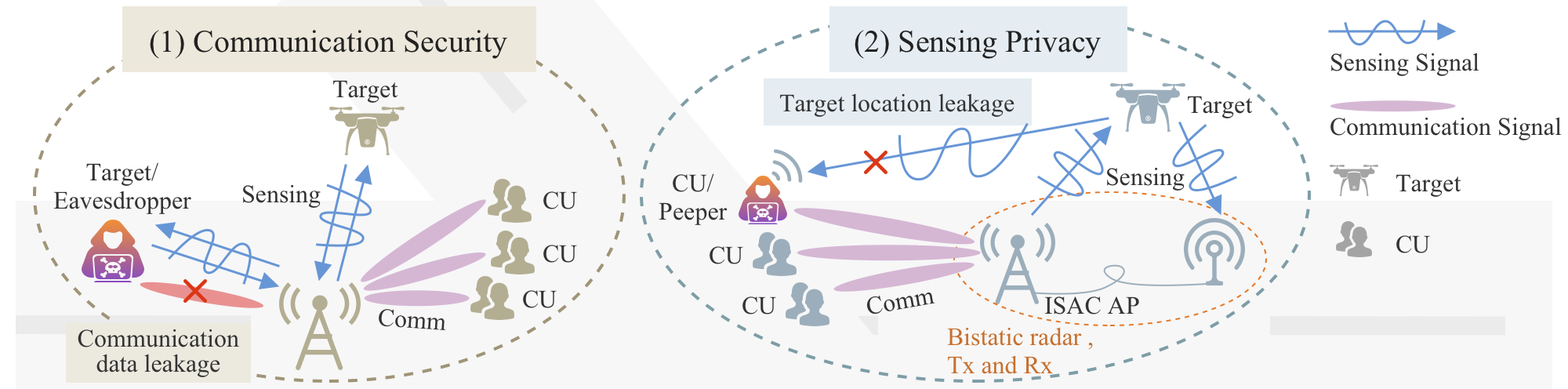}
        \caption{Communication Security and Sensing Privacy in ISAC Systems}
        \label{fig:SecurityandPrivacy}
    \end{figure*}

\section{ISAC Security and Privacy}\label{section5}

To fully explore the potential of ISAC technology in next-generation networks, the data security and sensing privacy should be properly addressed, as illustrated in Fig. \ref{fig:SecurityandPrivacy}. The integration of S\&C functions introduces new vulnerabilities due to dual-domain exposure. While sensing enhances communication, it may also inadvertently expose sensitive information, such as the location and movement patterns of targets. Furthermore, the shared infrastructure supporting both S\&C increases the risk of data interception, thereby heightening the likelihood of security breaches. Nonetheless, this integration also opens opportunities for designing enhanced security and privacy mechanisms\cite{yang2025privacy}.

\subsection{Communication Security}

Integrating communication data into probing ISAC signals introduces vulnerabilities, especially when eavesdroppers intercept these signals. The risk increases when high-power signals are used for target sensing. At the MAC layer, the shared data pathways between S\&C raise the possibility of routing integrity issues. Attackers may exploit sensing data for spoofing or impersonation, while the dual-use nature of ISAC amplifies the risk of man-in-the-middle attacks. At the physical layer, dual-functional beams used for both S\&C make communication data vulnerable to malicious eavesdropping, as targets in the beam’s path can intercept embedded communication signals. 

Physical layer security (PLS) remains vital in ISAC systems, even if communication data are encrypted at higher layers. PLS offers additional protection against low-level threats and strengthens overall system security by preventing attacks that could compromise communication channel data. Traditional secure beamforming approaches become insufficient because they may compromise sensing functionality. This necessitates the development of novel, tailored security approaches.

Several countermeasures to address communication data security in ISAC have been proposed.

\subsubsection{Artificial Noise (AN) Design}  
AN enhanced PLS by injecting interference to degrade the SINR at eavesdroppers, thus reducing their decoding ability while preserving communication and sensing quality for legitimate users. In \cite{9199556}, an optimization framework for AN-aided secure transmit beamforming in multi-user MISO ISAC systems maximized the secrecy rate by jointly designing beamformers and AN. It characterized the trade-offs between secrecy rate, communication quality, and sensing functionality and demonstrated robustness under imperfect S\&C CSI, confirming the effectiveness of AN in mitigating communication security threats.

\subsubsection{Directional Modulation (DM)}  
DM uses symbol-level precoding to apply constructive interference (CI) to legitimate receivers while limiting signals at potential eavesdroppers by employing destructive interference (DI). \cite{9737364} applied the CI-DI approach to mmWave frequencies, where angular correlations between S\&C channels are critical. This technique effectively enhances PLS by restricting signal leakage toward unauthorized directions, reducing energy consumption and system cost in high-frequency ISAC deployments.

\subsection{Sensing Privacy}

The integration of sensing functionality into wireless systems introduces additional privacy risks. Sensing signals emitted by dual-function BSs are reflected by targets. They may be intercepted by unauthorized receivers and exposing sensitive target-related information and compromising privacy. Unlike communication signals, which benefit from higher-layer encryption, passive sensing signals cannot be effectively encrypted, presenting a unique security challenge. Moreover, measures to mitigate interference between S\&C functions can inadvertently reveal critical transmitter information, such as location.  Enhanced sensing capabilities further elevate the risk of passive eavesdropping by unauthorized nodes. Additional concerns arise in imaging and environmental sensing, where unencrypted control signals (e.g., pilot signal) may be exploited to infer user positions or construct environmental maps. These vulnerabilities underscore the need for robust privacy-preserving mechanisms.

Several solutions have been proposed to mitigate sensing privacy risks.

\subsubsection{Beamforming for Privacy}  

Innovative beamforming techniques are crucial for preventing unauthorized data extraction while maintaining S\&C functionality. In \cite{10587082}, a beamforming-based method was proposed to secure sensing functionality by integrating AN. It optimized the beamforming and AN covariance matrix to enhance MI for legitimate radar users, while deteriorating it for unauthorized ones. This successfully restricts the adversary's ability to intercept sensing data, protecting environmental privacy without degrading sensing accuracy. Furthermore, \cite{10574259} proposed an AI-assisted adaptive beamforming method to obfuscate signals, blocking unauthorized sensing while keeping legitimate activities functional. This method suppresses unauthorized sensing while maintaining the effectiveness of legitimate sensing operations.

\subsubsection{CSI-Based User Location Anonymization}  
This method focuses on mitigating unauthorized positioning or tracking by obfuscating CSI changes. The goal is to obscure the user’s location while preserving sensor accuracy. In \cite{11004012}, an AI-based signal generation method is proposed to prevent unauthorized devices from extracting CSI-based location data, resulting in a 70\% reduction in the risk of inference. Similarly, \cite{10556606} proposed a multi-antenna signal-masking technique that exploits CSI time correlation and controls phase differences to thwart eavesdropping. Additionally, \cite{10611746} introduced a pilot-shielding method with tailored spectral characteristics to conceal genuine CSI variations, misleading eavesdroppers, and safeguarding location privacy.

\subsection{Sensing-Assisted Communication Security}

While the integration of sensing functionality in ISAC systems introduces new challenges in ensuring communication security and protecting sensor privacy, it also provides unique opportunities to enhance PLS. Leveraging its sensing capabilities, ISAC systems can dynamically estimate the sensing channel, i.e., the wiretap channel, by determining the angle of arrival of each target. This enables the system to prevent the target from decoding confidential information, thereby overcoming the traditional difficulty of obtaining eavesdropping CSI, a well-established limitation in conventional communication-only systems.

\subsubsection{Sensing-Assisted PLS}
Sensing-assisted PLS represents a promising approach for enhancing the security of ISAC systems, where the sensing functionality aids in improving the system’s ability to thwart eavesdropping and information leakage. By dynamically estimating the wiretap channel, ISAC systems can precisely track eavesdroppers’ locations and adapt communication strategies accordingly. This is particularly beneficial in overcoming the challenges posed by traditional methods that require difficult-to-obtain eavesdropping CSI. This allows for real-time estimation of eavesdropper angles, enabling secure beamforming strategies that degrade the signal at the eavesdropper’s location while boosting it for legitimate users. Such methods, as proposed in \cite{10349846} and \cite{11018844}, explore various optimization frameworks that balance communication performance with security objectives, demonstrating the mutual benefits of S\&C functionalities in enhancing PLS. In \cite{9838753}, the iterative optimization of beamforming and AN strategies results in improved secrecy rates by minimizing the potential for eavesdropping while dynamically adjusting to the eavesdropper’s estimated location. By combining S\&C capabilities, these systems represent a significant advancement in securing the wireless transmission of sensitive information in next-generation ISAC networks.


\subsubsection{Sensing-Assisted Covert Communication}


Unlike defending information eavesdropping in PLS, covert communication aims to prevent adversaries from detecting the existence of communication itself, requiring high-level confidentiality protection. Sensing-assisted covert communication leverages the dual functionality of radar and communication systems to ensure that information transmission remains undetectable by adversaries. These systems use sensing to track eavesdroppers and prevent communication signal detection by optimizing transmission characteristics, such as beamforming and power allocation. One approach involves jammer-assisted techniques, where a jammer is deployed to obscure transmission while the radar detects targets and communicates covertly. Optimization methods such as semi-definite relaxation (SDR) \cite{10507294} maximize covert throughput while maintaining radar detection constraints. Joint radar and communication beamforming designs also optimize covert throughput while maintaining radar performance. However, using dual-function waveforms introduces trade-offs between radar detection and covert transmission, balancing communication secrecy with efficient target detection. These approaches highlight ISAC’s potential to offer secure, covert communication by balancing the competing demands of S\&C performance.

\subsection{Lessons Learned}


Incorporating sensing data into security frameworks creates significant opportunities for co-design. However, several unresolved issues remain that warrant further investigation.

\subsubsection{How to Model and Defend Against Hybrid Cross-Domain
Attacks}

While extensive studies have investigated security and privacy in ISAC systems, few have systematically addressed hybrid threats, where adversaries simultaneously exploit vulnerabilities in both S\&C domains. Research on sensing-assisted eavesdropping \cite{10757770} highlighted this challenge, yet comprehensive attacker models that jointly capture sensing deception and communication interception remain scarce. Unlike conventional wireless security frameworks, ISAC requires hybrid security models that explicitly account for adversaries capable of manipulating both physical- and digital-layer information in a coordinated manner.


\subsubsection{How to Validate ISAC Security Frameworks under Real-World Constraints}

Most existing studies have relied on mathematically intensive optimization strategies, typically assuming perfect knowledge of attacker locations, CSI, and environmental conditions. In reality, ISAC systems must operate with imperfect CSI, mobility, hardware impairments, latency constraints, and limited sensing precision. 
Although \cite{10529955} investigated UAV-assisted ISAC security, its experimental validation remains limited, highlighting the need for prototypes, datasets, hardware testbeds, and standardized evaluation procedures.

\begin{figure*}[!htp]
    \centering
    \includegraphics[width=0.95\linewidth]{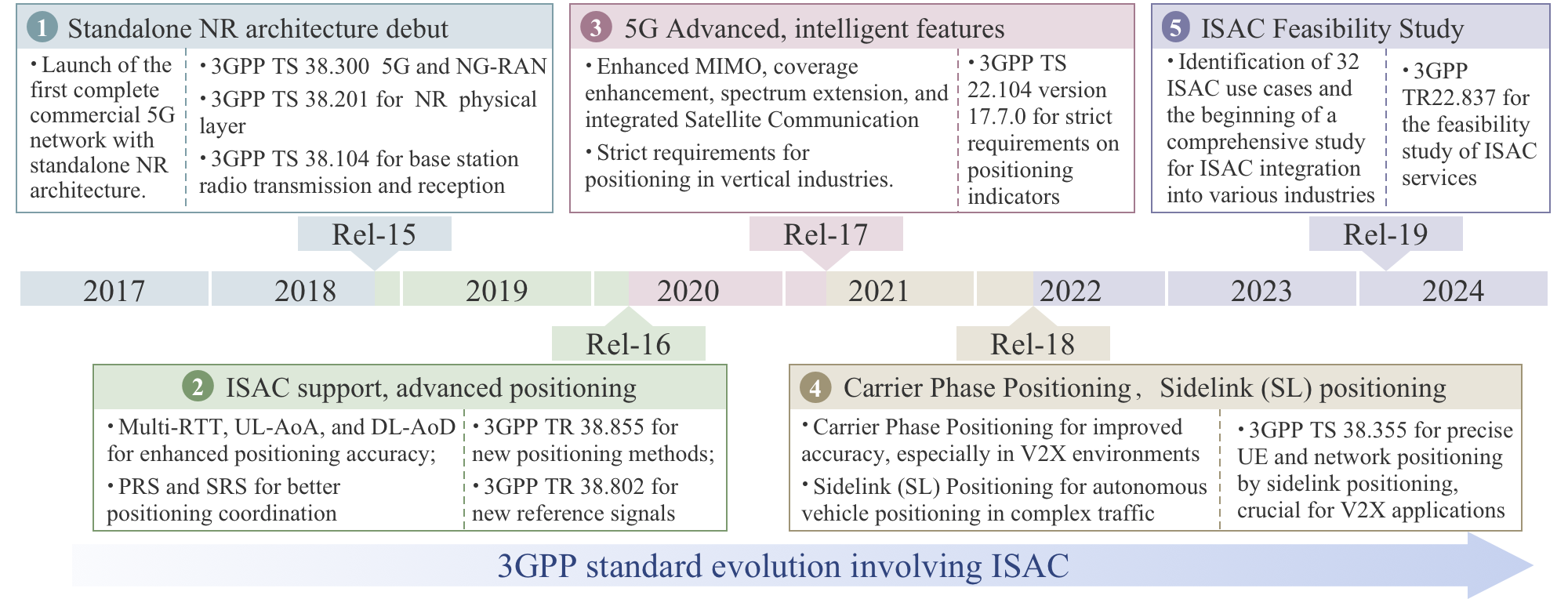}
    \caption{Brief Review of 3GPP Standards Progress in ISAC }
    \label{fig:3gpp}
\end{figure*}

\section{Standardization Progress from Third Generation Partnership Project (3GPP) to ITU}\label{section6}



Standardization has played a pivotal role in shaping the ISAC landscape, addressing interoperability challenges, enabling widespread adoption, and accelerating technological innovation. This section reviews recent progress in ISAC standardization, with particular focus on the collaborative roles of 3GPP, IEEE, and ITU. Key contributions on NR air interface advancements, emerging technologies, and regulatory frameworks are also examined.

\subsection{Efforts and Trends in Standardization}


The IEEE, alongside key organizations such as the ITU and 3GPP, has played complementary roles in the development and standardization of NR air interface technologies. 

\subsubsection{Brief Review of 3GPP Standards}


3GPP has progressively incorporated ISAC-enabling functionalities into its release-based standardization framework, thereby establishing the technical foundation for converged S\&C systems, as illustrated in Fig. \ref{fig:3gpp}.

\paragraph{Release 15: Foundational NR Technologies}
Rel-15 marked the launch of the first complete standalone NR architecture. Foundational technologies, including beamforming, massive MIMO, and flexible spectrum utilization, were introduced. These advancements, outlined in standards such as TS 38.300 and TS 38.104 \cite{TS38300, TS38104}, laid the groundwork for ISAC by enabling efficient resource allocation and high-performance signal processing. Although ISAC was not explicitly addressed, these technologies provided critical building blocks
for future integration.

\begin{figure}[htp!]
    \centering
    \captionsetup{justification=raggedright, singlelinecheck=false} 
    \includegraphics[width=0.99\linewidth]{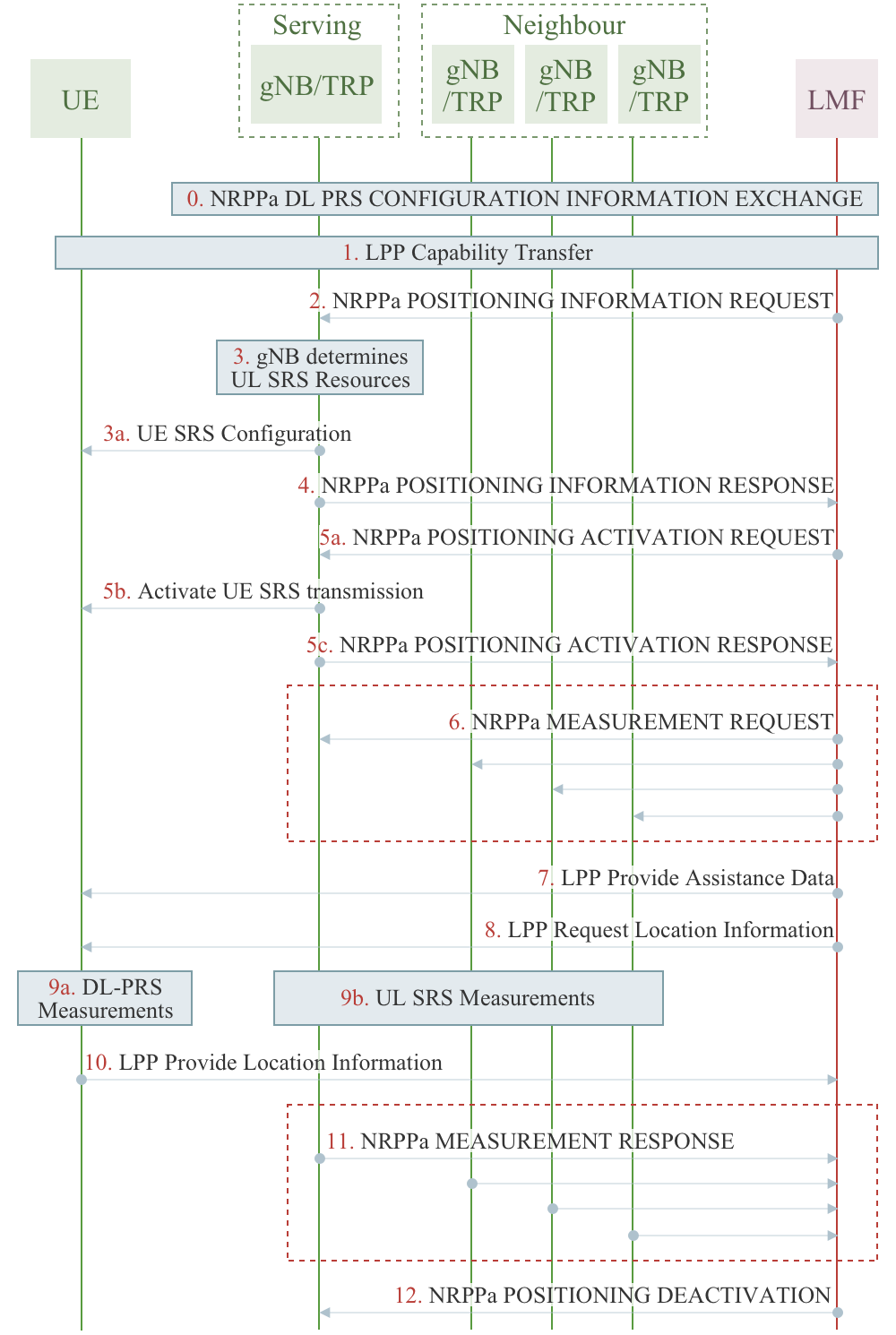}
    \caption{Multi-RTT positioning procedure: the messaging between the LMF, the gNBs, and the UE to perform the LMF-initiated location information transfer procedure for Multi-RTT.}
    \label{fig:Multi-rtt}
\end{figure}

\paragraph{Release 16: Positioning and Multicast for ISAC}
Rel-16 brought improved system efficiency, expanded use cases, and additional spectrum for 5G. Though it did not directly address ISAC, it introduced key enablers. Notably, 3GPP TR 38.855 introduced advanced positioning methods, including multi-round trip time (Multi-RTT), uplink-AoA, and downlink angle-of-departure (DL-AoD) \cite{TR38855}, improving accuracy in both indoor and outdoor environments. The positioning reference signal (PRS) and sounding reference signal (SRS) refined time-of-arrival and uplink-based positioning, with Fig. \ref{fig:Multi-rtt} showing the Multi-RTT procedure. Rel-16 also added multicast capabilities for vehicular use, enabling C-V2X communication. Initial mmWave support in Rel-16 paved the way for ISAC’s evolution in future 5G and 6G systems.

\paragraph{Release 17: mmWave ISAC Enhancements}
Rel-17 introduced advanced MIMO, improved coverage, spectrum extension, and satellite integration, forming the foundation for comprehensive 5G upgrades. In vertical industries like factory automation and transportation, 3GPP TS 22.104 imposed stringent positioning requirements, including 20–30 cm accuracy\cite{TS22104}. Rel-17’s 5G mmWave systems supported 400-MHz bandwidth, offering range estimation precision as fine as 0.1875 meters\cite{shi2022device}. Multi-antenna BSs exploit AoA/AoD estimation technologies such as multiple signal classification (MUSIC) for precise target detection. ISAC takes advantage of the extensive cellular infrastructure and high-frequency bands, reinforced by ultra-large-scale MIMO for increased bandwidth and superior angular resolution. By combining low-latency, high-precision communication with advanced sensing, ISAC facilitates remote object detection and environmental analysis without physical contact.



\paragraph{Release 18: 5G Advanced and Intelligent Positioning}

Rel-18 introduced 5G Advanced, featuring RAN intelligence and carrier-phase–based positioning, which significantly improved sensing accuracy. Key enhancements included sidelink positioning, enabling vehicles and devices to collaborate in localization in complex environments. The 3GPP TS 38.355 specification defined mechanisms for exchanging positioning information between UEs and networks, thereby enhancing V2X localization performance \cite{3gpp-ts-38.355,10517597}. This release also addressed multi-device coordination and dynamic resource allocation, laying the groundwork for more advanced ISAC integration in subsequent releases.

\paragraph{Release 19 and Beyond: Toward ISAC Standardization}

In early 2022, Rel-19 SA1 initiated a feasibility study on ISAC services, identifying use cases and requirements for commercial ISAC implementation across industries. By June 2023, the 3GPP SA plenary organized a workshop for Rel-19 SA2, where ISAC was proposed by several companies and included as a potential study item for Rel-19. TR 22.837 \cite{3GPP-rel19} identified 32 ISAC use cases across five deployment scenarios: smart transportation, smart industry, smart home, smart city, and smart drones.

In August 2023, 3GPP introduced TS 22.137, Service Requirements for ISAC, during the SA1\#103 meeting \cite{TS22137}. This specification defined eight KPIs for wireless sensing in 5G, including positioning accuracy, velocity estimation, and sensing resolution. Subsequently, in December 2023, the RAN\#102 meeting approved a dedicated study item on ISAC channel modeling, which was further refined at RAN\#103 in March 2024 \cite{R1-240}. The objective of this study is to establish a unified framework for object detection and tracking, thereby improving the reliability of ISAC across diverse scenarios such as UAVs, pedestrians, vehicles, and hazardous objects. Building upon these standardization efforts, ISAC air interface technologies are expected to be systematically investigated in Rel-20 beginning in 2025, with final specifications anticipated in Rel-21 as part of the 6G standardization cycle \cite{10918335}.

\paragraph{Summary and Discussion}

3GPP has made substantial progress in advancing converged S\&C technologies for advanced 5G systems. As NR technology continues to evolve, leveraging existing cellular infrastructure for perception functions offers clear benefits, including reduced deployment costs, accelerated rollout, and efficient spectrum reuse.


Enhancements in 5G infrastructure and air interface designs are expected to significantly strengthen wireless channels, thereby supporting more accurate and fine-grained environmental sensing. A representative example is Huawei’s proposal to integrate the sensing function into the 5G-Advanced core network, which aims to enable functions such as sensing control, AI-based data processing, secure data handling, and monetization of sensing data \cite{Huawei2023}. At the same time, the design of ISAC service processes and protocol stacks remains an open challenge. Interoperability requires well-defined signaling procedures for sensing capability negotiation, demand triggering, and result reporting. Whether existing protocol stacks can be extended to accommodate these new requirements, or if entirely new signaling frameworks must be developed, remains an open research problem. Addressing these issues will be essential to ensure seamless integration of ISAC into commercial deployments.

\subsubsection{Brief Review of IEEE Standards}

In parallel with 3GPP, IEEE has made significant strides in ISAC standardization. 
A major milestone is the IEEE 802.11bf standard, which incorporates sensing functionalities into the existing Wi-Fi ecosystem and contributes to the broader ISAC framework. Managed by the IEEE 802.11 Task Group bf, this effort aims to formalize Wi-Fi-based sensing within the IEEE 802.11 family, complementing 3GPP's efforts \cite{kaushik2024toward}.

\paragraph{Problem Identification and Framework Development}

In 2018, IEEE began exploring Wi-Fi sensing by leveraging CSI and the received signal strength indicator (RSSI) for environmental detection. Early studies demonstrated the feasibility of using reflected Wi-Fi signals to detect motion, presence, and physiological parameters without the need for additional hardware. Key use cases included indoor localization, smart home technologies, healthcare surveillance, and gesture recognition. The IEEE 802.11bf was instituted in 2020 to standardize Wi-Fi sensing. The initiative aimed to integrate sensing capabilities into existing Wi-Fi standards while ensuring backward compatibility and minimizing resource overhead so that communication performance is preserved. It also improved the ambiguity function through optimized waveform design and reflection analysis, leading to enhanced sensing accuracy.

\paragraph{Draft Development and Technical Enhancements}

By 2022, the IEEE 802.11bf draft standard had begun defining several foundational elements for Wi-Fi sensing. It formalized the use of existing Wi-Fi waveform components, including preamble training fields and pilot tones, to enable sensing without modifying the underlying frame structure, and it introduced sensing measurement reports to support the exchange of channel information for environmental inference. Refinements to OFDM-based waveforms improved sensing accuracy while preserving throughput, and enhancements to the ambiguity function enabled more reliable multi-target separation. The draft proposed mechanisms for interference mitigation, outlining dynamic strategies for allocating spectrum and time resources so that sensing operations and data transmission could coexist efficiently.

A comprehensive framework was defined, introducing dynamic resource allocation and optimized waveform designs to support dual S\&C functionalities.

\paragraph{Refinement and ISAC Integration}
The IEEE 802.11bf is refining the draft standards, with final approval anticipated in 2024. Efforts are concentrated on enhancing channel measurement techniques for precise location-based services and on aligning with adjacent fields such as autonomous systems and extended reality. In addition, multi-user resource management is being optimized to support dense deployment scenarios, which are critical for robust and scalable ISAC applications.

\paragraph{Summary and Discussion}
The IEEE 802.11bf standard has substantially advanced ISAC standardization by exploiting existing Wi-Fi features for dual-use S\&C. Dynamic resource allocation mitigated interference challenges, ensuring reliable performance for both functions. Moreover, multi-device collaboration frameworks extend sensing coverage and accuracy, laying the foundation for scalable ISAC solutions in future Wi-Fi and beyond systems.

\subsubsection{Brief Review of ITU Standards}

In parallel with the efforts of 3GPP and IEEE, ITU has primarily focused on spectrum management and regulatory alignment, with particular emphasis on spectrum allocation and the development of frameworks that facilitate ISAC deployment in higher-frequency bands.

\paragraph{Spectrum Allocation and Management}

At the World Radiocommunication Conference 2019 (WRC-19), ITU allocated higher frequency bands, including mmWave spectrum in the 24--52.6 GHz range, to support 5G. These allocations are highly favorable for ISAC, as the wide bandwidth and short wavelengths enable fine sensing resolution. In the next phase, ITU has also initiated studies on sub-terahertz (sub-THz) and terahertz (THz) bands (100-300 GHz), which are expected to be critical for high-resolution S\&C in 6G networks \cite{kaushik2024toward}.

\paragraph{IMT Framework and ISAC Integration}

ISAC has been formally incorporated into the IMT-2030 framework as part of ITU’s 6G vision. This framework focuses on dynamic spectrum sharing and radar–communication coexistence, enabling advanced ISAC deployments.

\paragraph{Summary and Discussion}
ITU’s contributions center on spectrum allocation, global harmonization, and aligning standards with ISAC objectives. Prioritizing smart spectrum coexistence and dynamic sharing, ITU advanced frameworks support ISAC integration. ITU has also initiated discussions on the role of metasurfaces and RIS, highlighting their potential to enhance sensing resolution and optimize the performance of ISAC systems.

\subsection{Lessons Learned}


Standardization has advanced ISAC from a conceptual vision to a pre-commercial reality, yet several technical and organizational challenges remain. Insights from current standardization efforts, aligned with emerging research demands, point to the following key questions:

\subsubsection{How Can ISAC Performance be Validated through Standardized KPIs}

Standardization bodies have defined performance metrics and testing methodologies to enable systematic evaluation of ISAC. IEEE 802.11bf specifies sensing KPIs such as velocity (0.1–0.4 m/s) and angular resolution (1°–8°), which are critical for assessing the trade-offs between S\&C performance. 3GPP further incorporates these KPIs into channel models to validate ISAC under real-world cellular environments, particularly in vehicular networks and IoT systems \cite{10251772}. These benchmarks ensure that standardized ISAC solutions achieve acceptable trade-offs between sensing accuracy and communication reliability.


\begin{figure}
    \centering
    \includegraphics[width=0.9\linewidth]{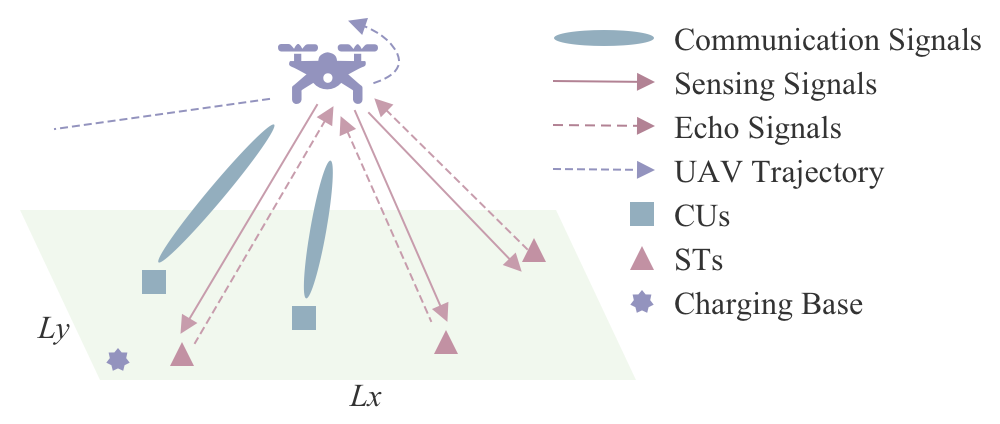}
    \caption{ISAC-based rotary-wing UAV Scenario \cite{10529184}}
    \label{fig:uav1}
\end{figure}

\subsubsection{How Can Use Case-Driven Research Guide Standardization} 


3GPP SA1 has consolidated extensive research on ISAC scenarios within the RAN1 channel-modeling project, with industry consensus highlighting the pivotal roles of the low-altitude economy, drones, and intelligent transportation. For example, enhancements in 5G NR Rel-18 enable drones to support beamforming, allowing BSs to detect unauthorized flights, perform obstacle avoidance, and guide trajectories via multi-station collaboration. Integrating passive tracking with active sensing improves positioning accuracy, while environmental awareness assists beam management and mobility. Furthermore, mmWave operation achieves low latency with slot lengths as short as 0.125 ms, ensuring real-time responsiveness.


The low-altitude economy has also gained strong policy support. In China, the Civil Aviation Administration has promoted 5G NR–based applications, and the Ministry of Industry and Information Technology announced on October 23 initiatives to establish ISAC-enabled infrastructure for low-altitude networks\footnote{\url{https://english.www.gov.cn/news/202410/23/content_WS6718e264c6d0868f4e8ec381.html}}. Despite these advances, practical deployments face challenges such as dense connectivity, interference, and spectrum conflicts. To address these, joint optimization methods have been explored, including beamforming, dynamic spectrum allocation, and coordinated frequency management. For instance, \cite{lyu2022joint} proposed a joint UAV maneuver and beamforming strategy that improved trajectory planning, bandwidth allocation, and overall energy efficiency. These optimization techniques significantly enhance system performance, as demonstrated in \cite{10529184} and Fig.~\ref{fig:uav1}.


\section{Conclusion and Future Directions}\label{section7}

This paper has extensively examined the impact of ISAC across key domains, including operating spectrum, network architecture, sensing methodologies, security challenges, and standardization efforts within the context of 5G and beyond. These developments provide insights into ISAC’s potential and the opportunities it offers across multiple applications.


ISAC-driven advancements have already demonstrated substantial benefits across diverse domains. In intelligent transportation, it delivers real-time road condition data to enhance autonomous driving safety and efficiency. In smart cities, it supports adaptive infrastructure that dynamically responds to residents’ needs. In industrial IoT, it facilitates high-precision monitoring and predictive maintenance, improving operational efficiency and reducing failures. These innovations not only improve existing systems but also enable entirely new services that were previously unattainable.

As ISAC matures, its role in wireless communication will likely expand, shaping more efficient, intelligent, and adaptive networks. 
By embedding sensing functions into the communication infrastructure, ISAC provides wireless systems with the ability to understand and respond to the environment in real time. However, this evolution has also brought many challenges, including complex resource management mechanisms, higher-intensity security requirements, and scalable hardware integration schemes.

Building on current progress, several open research directions can be identified:



\begin{itemize}
    \item Unified channel modeling: 
    Unlike traditional communication models, ISAC demands hybrid geometric-stochastic representations that take into account multipath reflections, clutter, and scattering caused by objects. Although the continuous research of 3GPP RAN1 \cite{R1-240} has taken an important step, future standards still need to adopt multi-bounce, clutter aware and object-tagged models to support sub-THz and mmWave sensing with high fidelity.

     \item  Resource optimization under multi-objective constraints:  Emerging technologies such as rate-splitting multiple access (RSMA) \cite{xu2021rate} and OTFS waveform design \cite{gaudio2020effectiveness} show good prospects, but the existing scheduling framework rarely incorporates KPIs specific to sensing, such as angle resolution or Doppler sensitivity. Developing ISAC-aware scheduling policies that dynamically balance spatial, temporal, and spectral resources represents a compelling direction for ongoing research.


     
     \item Privacy and Security: The exposure of environmental and user-related data introduces significant risks. Beyond conventional encryption, sensing-aware systems require native security features, such as federated access control, encrypted beamforming, and zero-trust architectures \cite{wei2022toward}. At the same time, these mechanisms must align with evolving global data protection regulations to ensure lawful and ethical deployment.

\end{itemize}

In conclusion, ISAC is poised to become a cornerstone of future wireless networks, unifying spectrum utilization, network intelligence, and sensing awareness into a coherent design paradigm. Its successful realization will depend on advancing unified models, resource-aware architectures, and privacy-preserving frameworks. Ongoing interdisciplinary research will be indispensable for translating ISAC’s conceptual promise into scalable, real-world deployments, ultimately shaping the foundations of 6G and beyond. The true measure of ISAC’s success will lie not only in its technical feasibility but also in its ability to foster new applications, enhance resilience, and generate societal value across next-generation wireless ecosystems.

\ifCLASSOPTIONcaptionsoff
  \newpage
\fi

{
\small
\bibliographystyle{IEEEtran}
\bibliography{bibtex/bib/IEEEabrv,bibtex/bib/IEEEexample}
}
\end{document}